\documentclass[lettersize,journal]{IEEEtran}
\usepackage{amsmath,amsfonts}
\usepackage[linesnumbered,ruled,vlined]{algorithm2e}

\usepackage{array}
\usepackage[caption=false,font=footnotesize]{subfig}
\usepackage{textcomp}
\usepackage{stfloats}
\usepackage{url}
\usepackage{verbatim}
\usepackage{amssymb} 
\usepackage{amsfonts}
\usepackage{amsmath} 
\usepackage{physics} 

\usepackage{color} 
\usepackage{bm}
\usepackage{epstopdf} 
\usepackage{graphicx}
\usepackage{booktabs}  
\usepackage[dvipsnames]{xcolor} 
\usepackage{cite} 
\usepackage{balance}
\usepackage{setspace}  
\usepackage{amsthm}
\usepackage{amsmath} 
\usepackage{hyperref}
\usepackage{balance}
\usepackage{placeins} 
\usepackage{scalerel}

\allowdisplaybreaks 

\newtheorem{remark}{Remark}

\newtheorem*{pf}{Proof}  



\newcommand{\figref}[1]{Fig. \ref{#1}}
\newcommand{\secref}[1]{Section \ref{#1}}

\definecolor{softred}{RGB}{200, 50, 50}

\DeclareMathAlphabet{\mathsfit}{\encodingdefault}{\sfdefault}{m}{sl}
\SetMathAlphabet{\mathsfit}{bold}{\encodingdefault}{\sfdefault}{bx}{n}
\newcommand{\tens}[1]{\bm{\mathsfit{#1}}}

\def\tI{{\tens{I}}}

\def\tQ{{\tens{Q}}}

\def\tX{{\tens{X}}}

\begin{document}

\bstctlcite{IEEEexample:BSTcontrol}

\title{Semantic Communication for 
\\Multi-Satellite Massive MIMO Transmission: \\
A Mixture of Cooperative Modes Framework}    
	
\author{Yafei Wang, \textit{Graduate Student Member}, \textit{IEEE}, Yuchen Zhang, \textit{Member}, \textit{IEEE}, 
\\ Yiming Zhu, \textit{Graduate Student Member}, Vu Nguyen Ha, \textit{Senior Member}, \textit{IEEE}, Rui Ding,\\ Wenjin Wang, \textit{Member}, \textit{IEEE},
Symeon Chatzinotas, \textit{Fellow}, \textit{IEEE}, Björn Ottersten, \textit{Fellow}, \textit{IEEE}
\thanks{Yafei Wang, Yiming Zhu, and Wenjin Wang are with the National Mobile Communications Research Laboratory, Southeast University, Nanjing 210096, China, and also with Purple Mountain Laboratories, Nanjing 211100, China (E-mail: \{wangyf,  ymzhu, wangwj\}@seu.edu.cn).}
\thanks{Yuchen Zhang is with the Computer, Electrical, and Mathematical Science \& Engineering (CEMSE) Division, King Abdullah University of Science and Technology (KAUST), Thuwal 23955-6900, Kingdom of Saudi Arabia (E-mail: yuchen.zhang@kaust.edu.sa).}
\thanks{Rui Ding is with China Satellite Network Group Company Ltd., Beijing
100029, China (E-mail: greatdn@qq.com).}
\thanks{Vu Nguyen Ha, Symeon Chatzinotas, and Björn Ottersten are with the Interdisciplinary Centre for Security, Reliability and Trust (SnT), University of Luxembourg (E-mails: \{vu-nguyen.ha, symeon.chatzinotas, bjorn.ottersten\}@uni.lu).}
}

\markboth{}%
{Shell \MakeLowercase{\textit{et al.}}: A Sample Article Using IEEEtran.cls for IEEE Journals}

\maketitle	
	
\IEEEpeerreviewmaketitle

\begin{abstract}

This paper investigates semantic communications (SemComs) for multi-satellite cooperative massive multiple-input multiple-output (MIMO) transmission, where multiple massive-MIMO satellites jointly serve a common set of multi-antenna user terminals. For the first time, SemComs with image transmission task are integrated into satellite massive MIMO and multi-satellite cooperative transmission. For the two representative cooperative modes, namely coherent transmission (CT) and non-coherent transmission (NCT), we develop multi-satellite CT (MSCT) and multi-satellite NCT (MSNCT) SemCom frameworks, respectively. MSCT adopts a symmetric architecture, whereas MSNCT introduces transmitter-side stream allocation and a two-stage receiver design that combines per-stream semantic extraction with cross-stream semantic-interference exploitation. 
To instantiate MSCT, we further design a symmetric encoder and decoder network based on hybrid Swin-Transformer and lightweight bottleneck convolutional neural network (CNN) blocks, termed HSTC, where Swin Transformer provides scalable computation and the CNN branch improves performance and convergence. 
For MSNCT, a Transformer-based backbone is employed to support cross-stream interference exploitation through global attention. Building on these two frameworks, we propose a mixture of cooperative modes (MoCM) framework, in which a permutation-invariant network dynamically switches between MSCT and MSNCT using multi-satellite statistical channel state information, thereby balancing semantic performance and complexity. Simulation results under practical configurations demonstrate the performance gains of the proposed frameworks.
\end{abstract}
\vspace{-3mm}


\IEEEpeerreviewmaketitle
	
    \vspace{-3mm}	
        
    \section{Introduction}
    
    \vspace{-1mm}
    
        \IEEEPARstart{S}{ixth}-generation (6G) wireless networks regard satellite communications (SatComs) as an indispensable component of ubiquitous connectivity, since satellite networks can effectively complement the coverage limitations of terrestrial cellular systems in oceans, deserts, mountainous regions, and aerial spaces \cite{10820534,WANG2025}. To further improve spectral efficiency and system capacity, introducing massive multiple-input multiple-output (MIMO) transmission into SatCom is of great importance, as the additional spatial degrees of freedom can enhance both spatial multiplexing and link gain, thereby better supporting future high-density and heterogeneous access demands \cite{wu2024large, ASTBlueWalker3}. Nevertheless, the long propagation delay and pronounced Doppler dynamics in SatCom make the acquisition and feedback of instantaneous channel state information (CSI) highly challenging. By contrast, statistical CSI (sCSI), owing to its lower acquisition overhead, slower variation, and stronger robustness, provides a more practical basis for transmission design in satellite massive MIMO systems \cite{wang2026DP_JSAC}.

        With the development of mega-constellations and inter-satellite link (ISL) capabilities, multi-satellite cooperative transmission offers a new technical path for improving link reliability, system capacity, and service continuity by jointly exploiting the spatial, power, and coverage resources of multiple satellites \cite{wang2026stf,wang2026DP_JSAC,xiang2024massive}. Meanwhile, semantic communications (SemComs), which emphasize task effectiveness and semantic fidelity, are better suited to improving effective transmission efficiency under limited communication resources, making them particularly attractive for satellite scenarios with tight link budgets and low-signal-to-noise ratio (SNR) operating conditions \cite{Xie2021DeepSC,SemanticCommSurvey,Lin2026SatelliteSemCom}. If SemComs are matched to the specific transmission mechanisms of multi-satellite cooperation, the link-budget enhancement and channel-rank improvement brought by multi-satellite cooperation can mutually reinforce the task-oriented advantages of SemComs, yielding synergistic gains beyond simple superposition. Guided by this rationale, this paper develops SemCom frameworks tailored to coherent transmission (CT) with a common data stream and non-coherent transmission (NCT) with distinct data streams, respectively, and further designs a mixture of cooperative modes (MoCM) framework, thereby forming an innovative multi-satellite SemCom system.
    \vspace{-3mm}
    
    \subsection{Related Works}   

    Previous studies have extensively investigated the channel characteristics and downlink transmission design of satellite massive multiple-input multiple-output (MIMO) systems \cite{wu2024large}. In particular, \cite{you2020massive} analyzed the channel properties of satellite massive MIMO and proposed an sCSI-based downlink transmission scheme, while \cite{li2021downlink} further studied downlink transmit design for satellite massive MIMO and showed that exploiting slowly varying sCSI can provide an effective tradeoff between performance and signaling overhead. In addition, subsequent works extended this line of research toward transmission design and architectural evolution \cite{10437228,you2022hybrid}. Nevertheless, the communication capability of a single satellite remains fundamentally limited by payload constraints, array aperture, and link budget. With the growth of mega-constellations and the enhancement of inter-satellite interaction capabilities, multi-satellite cooperative transmission has attracted increasing attention because it can jointly exploit the spatial \cite{wang2026DP_JSAC, ha2024user}, temporal \cite{11449148}, and frequency \cite{9193995} resources of multiple satellites to improve coverage continuity, link reliability, and system capacity \cite{wang2026stf,Bakhsh2024MultiSatSurvey,wang2025MSMS}. In terms of transmission mechanisms, existing spatial-domain multi-satellite cooperation mainly includes CT and NCT \cite{wang2026stf}. In the former, multiple satellites deliver the common data stream to the user terminal (UT) and achieve higher link budget and stronger coverage gain through coherent combining, making it more suitable for UTs with a small number of receive antennas \cite{wang2026DP_JSAC,wu2025distributed,zhang2026enabling,zhang2025decentralized}.
    In the latter, different satellites transmit distinct streams to the same UTs, increasing stream multiplexing at the cost of stronger cross-stream interference and thus requiring more capable multi-antenna receivers \cite{xiang2024massive,cao2026DLMSCT}.



    SemComs have recently emerged as a new paradigm distinct from conventional bit-level transmission, with the primary goal of preserving task effectiveness and semantic fidelity under limited communication resources. For example, \cite{Xie2021DeepSC} developed a deep learning (DL)-enabled end-to-end SemCom framework that optimizes semantic similarity instead of exact bit recovery, thereby demonstrating the potential of semantic-level transmission. In addition, \cite{DeepJSCCMIMO2024} investigated adaptive image transmission over MIMO channels and proposed a Vision Transformer (ViT)-based DeepJSCC-MIMO architecture. Overall, SemComs offer notable robustness and efficiency in resource-limited regimes, which is particularly attractive for satellite systems operating under tight link budgets \cite{SemanticCommSurvey}. Moreover, their typical task types oriented toward images and speech are also highly aligned with the requirements of SatCom scenarios \cite{Liu2026SatAVSemCom,Bui2025LEOEO,helber2019eurosat}. In particular, \cite{Lin2026SatelliteSemCom} studied robust semantic transmission for SatComs, while \cite{JSCCSatGroundSemCom} investigated multimodal joint source-channel coding for satellite-to-ground SemComs. Nevertheless, these works are still mainly confined to single-satellite-to-UT links, and do not further investigate SemCom mechanisms tailored to multi-satellite cooperative modes, e.g., CT and NCT, which exhibit distinct link characteristics. On the other hand, distributed SemComs have also begun to attract attention. Representative studies investigated image SemCom over shared wireless channels \cite{DeepMA2024}, as well as image-delivery optimization in cooperative SemCom networks \cite{Zhang2024CoopImageSemCom}. 
    However, existing methods have not fully exploited the system characteristics of specific communication scenarios, nor have they sufficiently explored cooperative transmission modes.
    
    \vspace{-2mm}
    \subsection{Contributions}
    \vspace{-1mm}
        To support more efficient and intelligent connectivity in wide-area 6G scenarios, it is necessary to develop SemCom schemes that are well matched to multi-satellite cooperative transmission mechanisms. On the one hand, SemComs can improve effective semantic transmission efficiency under the power-constrained conditions of satellite systems \cite{Xie2021DeepSC,DeepJSCCMIMO2024,SemanticCommSurvey}. On the other hand, multi-satellite cooperative transmission can further enhance system capacity by jointly exploiting the spatial resources of multiple satellites \cite{wang2026stf,wang2026DP_JSAC,wang2025MSMS}. 
        A deep and synergistic integration of the two is therefore expected to fully exploit their complementary strengths. However, existing studies on multi-satellite cooperative transmission have not incorporated SemComs, leaving the cooperative gains, the characteristics of massive MIMO, and multiple cooperative modes underexplored within a SemCom framework. This leads to a key research question: \textit{How to design a SemCom framework for multi-satellite massive MIMO transmission?} This paper answers this question by developing a MoCM framework for multi-satellite cooperative SemCom. The major contributions of this work are as follows:
        \vspace{-3.5mm}
        \begin{itemize}
            \item For multi-satellite cooperative massive MIMO SemComs, we establish a multi-satellite downlink signal model with sCSI. Building on this model, we develop the multi-satellite CT (MSCT) SemCom framework, where multiple satellites transmit a common semantic stream through parameter-shared semantic encoders, while the receiver constructs a low-complexity joint receive beam based on sCSI.   The effective channel and noise features are provided to the semantic decoder as auxiliary conditions. Moreover, we design a symmetric hierarchical encoder-decoder network built upon Hybrid Swin-Transformer and CNN (HSTC) blocks, where window attention preserves scalable complexity and the lightweight bottleneck CNN branch complements local semantic modeling without introducing excessive parameter overhead.
            \item For NCT, where different satellites transmit distinct and mutually interfering semantic streams, we develop a tailored multi-satellite NCT (MSNCT) framework. It first partitions the source semantic information into multiple sub-semantic streams at the transmitter, which are then delivered by satellites through parameter-shared semantic encoders. We further propose a two-stage receiver-side semantic decoding architecture, where the first stage recovers stream-wise semantic features through a parameter-shared extraction network, while the second stage performs cross-stream semantic interference exploitation in the token domain. The Transformer-based backbone enables effective cross-stream semantic interaction through global self-attention. 
            \item
            To adaptively exploit the advantages of MSCT and MSNCT under different channel conditions and complexity requirements, we further propose the MoCM framework. The framework dynamically selects between MSCT and MSNCT according to the tradeoff between semantic reconstruction performance and complexity. To enable low-overhead and scalable mode selection, we design a low-dimensional input representation based on long-term multi-satellite sCSI and construct an attention-based permutation-invariant mode-switching network. The network realizes multi-satellite CSI interaction through satellite-dimension Transformer blocks, extracts global features via attention-based pooling, and finally selects the cooperative mode.
            \item 
            Based on practical low-Earth-orbit (LEO) constellation settings and Monte Carlo geometry simulations, we comprehensively evaluate the proposed frameworks and networks under different transmit powers, receive-antenna configurations, and compression ratios. The results highlight the stable gains of the multi-satellite cooperative semantic frameworks, the effectiveness of cross-stream semantic-interference exploitation in MSNCT, and the ability of MoCM to adaptively deliver the best performance under a favorable complexity tradeoff.
        \end{itemize}

    
    The remainder of this paper is organized as follows. \secref{sec:system_model} introduces the system model and the received signal model for multi-satellite SemComs. \secref{framework ct sec} and \secref{framework nct sec} present the SemCom frameworks and network designs for MSCT and MSNCT, respectively. \secref{sec:scalable_csi_module} develops the MoCM framework. \secref{sec:simulation_results} provides simulation results, and \secref{sec:conclusion} concludes this paper.
    
    {\textit{Notation}}: $x$, ${\bf x}$, ${\bf X}$, and $\tX$ denote a scalar, vector, matrix, and tensor, respectively. $(\cdot)^T$, $(\cdot)^{*}$, $(\cdot)^H$, and $(\cdot)^{-1}$ denote transpose, conjugate, conjugate transpose, and inverse, respectively. $\mathbb{R}$ and $\mathbb{C}$ denote the real and complex domains. $\mathbb{E}\{\cdot\}$, $\|\cdot\|_2$, $\otimes$, and ${\rm Tr}(\cdot)$ denote expectation, the Euclidean norm, the Kronecker product, and the matrix trace, respectively. ${\bf I}_{M}$ is the $M\times M$ identity matrix, $|\mathcal{A}|$ is the cardinality of set $\mathcal{A}$, and $\mathcal{CN}(\mu,\sigma^2)$ denotes a circularly symmetric complex Gaussian distribution.

\vspace{-2mm}
\section{System Model}
\begin{figure}[t]
    \centering
    \includegraphics[width=0.9\columnwidth]{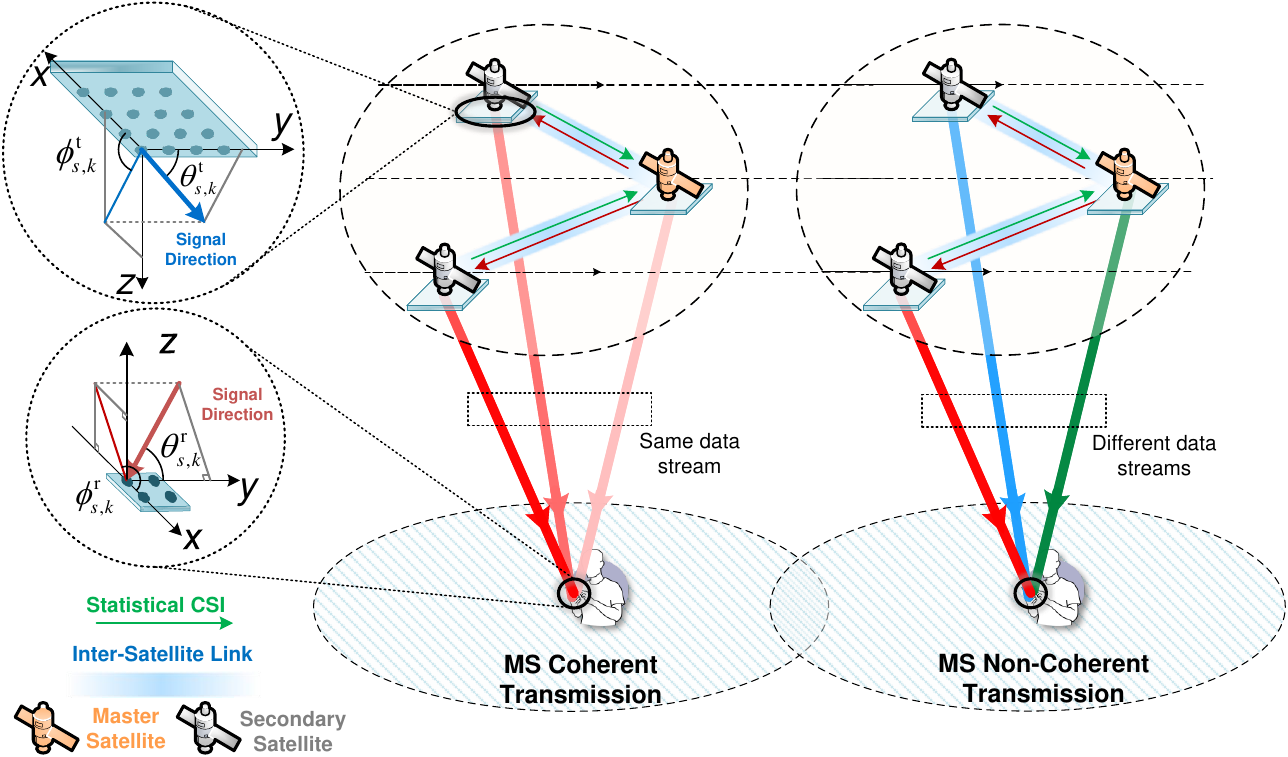}
    \caption{Cooperative multi-satellite systems with CT and NCT.}
    \label{system model fig}
    \vspace{-5mm}
\end{figure}

\vspace{-1mm}
\label{sec:system_model}
As shown in \figref{system model fig}, we consider the downlink of a multi-satellite system, where multiple UTs within a coverage area are served by $S$ satellites.
The inter-user interference is assumed to be eliminated via frequency division, so that we can focus on the per-UT transmission design.
Without loss of generality, satellites and users employ uniform planar arrays (UPAs) with $N_{\rm T}=N_{\rm TV}N_{\rm TH}$ and $N_{\rm R}=N_{\rm RV}N_{\rm RH}$ antennas, respectively. 
The source information for downlink transmission is assumed to be available within the serving satellite cluster. 
As illustrated in the tables of \figref{system model fig}, each UT is served by multiple satellites, which constitute the satellite set $\mathcal{S}_k$.

\vspace{-3mm}	
    \subsection{Multi-Satellite Channel Model}
    \label{channel sec}
    \vspace{-1mm}	
    
    According to \cite{li2021downlink,hou2024joint,zhu2024joint}, the time-varying spatial domain MIMO channel ${\tilde {\bf H}}_{s,k}(t, \tau)\in\mathbb{C}^{N_{\rm R}\times N_{\rm T}}$ between the $s$-th satellite and the $k$-th UT is given by
    \vspace{-2mm}
    \begin{align}
    {\tilde {\bf H}}_{s,k}(t, \tau) \!=\! \sum_{l=1}^{L_{s,k}}a_{s,k,l}\cdot{\rm e}^{j2\pi \nu_{s,k,l}t}\delta(\tau\!-\!\tau_{s,k,l}){\bf u}_{s,k,l}{\bf v}^T_{s,k},
    \label{delay channel eq}
    \end{align}
    where $t$ and $\tau$ represent time and delay; $l$ is the index of the path, and the total number of paths in the channel is $ L_{s, k} $; $a_{s,k,l}$, $\nu_{s,k,l}$, and $\tau_{s,k,l}$ represent the complex gain, Doppler frequency offset, and delay of the path channel, respectively; ${\bf u}_{s,k,l}\in\mathbb{C}^{N_{\rm R}\times 1}$ and ${\bf v}_{s,k}\in\mathbb{C}^{N_{\rm T}\times 1}$ are the steering vectors at the receiver and transmitter.

    For multi-satellite cooperative transmission, we assume that the satellites perform satellite-side pre-compensation for delay, Doppler, and phase, so that their signals can be effectively aligned and superposed at the receiver \cite{zhu2026multisatellitecooperative, wang2019near,marrero2022architectures}.
    Then, the received signal of UT $k$ over the considered time-frequency resource can be written as
    \vspace{-2mm}
    \begin{align}
     \textstyle{\bf y}_{k} \!=\! \sum_{s\in\mathcal{S}_{k}}\!{\bf H}_{s,k} {\bf x}_{s,k} +  {\bf n}_{k},
     \label{received signal eq1}
    \end{align}
    where ${\bf x}_{s, k}\in\mathbb{C}^{N_{\rm T}\times 1}$ is the precoded desired signal transmitted from satellite $s$ to UT $k$. ${\bf n}_k\in\mathbb{C}^{N_{\rm R}\times 1}$ denotes the additive white Gaussian noise vector with distribution ${\mathcal{CN}}({\bf 0}, \sigma^2_k{\bf I})$. 
    In \eqref{received signal eq1}, ${\bf H}_{s, k}\in\mathbb{C}^{N_{\rm R}\times N_{\rm T}}$ denotes the channel frequency response dominated by the line-of-sight (LoS) path after pre-compensation, expressed as \cite{li2021downlink}, \cite{wu2025distributed}
\begin{align}
    {\bf H}_{s,k} &\textstyle= \sqrt{\frac{\kappa_{s,k}\gamma_{s,k}}{\kappa_{s,k} + 1}}{\bf H}^{\rm LoS}_{s,k} + \sqrt{\frac{\gamma_{s,k}}{\kappa_{s,k} + 1}}{\bf H}^{\rm NLoS}_{s,k}\\
    &\textstyle= \left(\sqrt{\frac{\kappa_{s,k}\gamma_{s,k}}{\kappa_{s,k} + 1}}{\bf u}_{s,k} + \sqrt{\frac{\gamma_{s,k}}{\kappa_{s,k} + 1}}{\tilde {\bf u}}_{s,k}\right){\bf v}^T_{s,k}, \label{frequency channel model eq}
\end{align}
    where $\gamma_{s,k} = {\mathbb E}\{{\rm Tr}({\bf H}_{s,k}{\bf H}^H_{s,k})\}$ represents the average channel power, and $\kappa_{s,k}$ denotes the Rician factor. ${\bf H}^{\rm NLoS}_{s,k}={\tilde {\bf u}}_{s,k}{\bf v}^T_{s,k}$ is the random non-line-of-sight (NLoS) channel introduced by scatterers around the UT, characterizing the NLoS component in \eqref{delay channel eq}, where ${\tilde {\bf u}}_{s,k}\sim{\mathcal {CN}}({\bf 0}, {\boldsymbol{\Sigma}}_{s,k})$. ${\bf H}^{\rm LoS}_{s,k}={\bf u}_{s,k}{\bf v}^T_{s,k}$ denotes the dominant LoS path channel, where the expressions for the steering vectors ${\bf v}_{s,k}$ and ${\bf u}_{s,k}\triangleq {\bf u}_{s,k,1}$ are expressed as
    \vspace{-2mm}
    \begin{align}
        {\bf v}_{s,k} &= {\bf v}_{N_{\rm TV}}(\cos ({\theta^{\rm t}_{s,k}}))\otimes  {\bf v}_{N_{\rm TH}}(\sin(\smash{\theta^{\rm t}_{s,k}})\cos(\phi^{\rm t}_{s,k})),\label{eq v}\\
        {\bf u}_{s,k} &= {\bf v}_{N_{\rm RV}}(\cos ({\theta^{\rm r}_{s,k}}))\otimes  {\bf v}_{N_{\rm RH}}(\sin(\smash{\theta^{\rm r}_{s,k}})\cos(\phi^{\rm r}_{s,k})),\label{eq u}
    \end{align}
    where ${\phi^{\rm t}_{s,k}}$ and ${\theta^{\rm t}_{s,k}}$ are the departure azimuth and elevation angles of the signal, respectively; ${\phi^{\rm r}_{s,k}}$ and ${\theta^{\rm r}_{s,k}}$ are the arrival azimuth and elevation angles of the signal, with their specific definitions illustrated in \figref{system model fig}. For simplicity of notation, we define ${\boldsymbol{\theta}}_{s,k}\triangleq[{\theta^{\rm t}_{s,k}},{\phi^{\rm t}_{s,k}}, {\theta^{\rm r}_{s,k}},{\phi^{\rm r}_{s,k}}]^T$. The vector ${\bf v}_{N}(x)\in\mathbb{C}^{N\times 1}$ is defined as $\textstyle {\bf v}_{N}(x) = \frac{1}{\sqrt{N}}\cdot [{\rm e}^{-j\pi0x}, {\rm e}^{-j\pi1x}, ..., {\rm e}^{-j\pi(N-1)x}]^T$.

    While each LoS-dominant satellite-to-UT link supports only a single data stream, a multi-antenna UT enables different satellites to transmit distinct streams simultaneously \cite{you2020massive,xiang2024massive}.
    In addition, real-time estimation of the instantaneous massive MIMO channel ${\bf H}_{s,k}$ is challenging because of satellite mobility. The LoS-dominant satellite channel instead makes an sCSI-based framework feasible. 
As shown in \figref{system model fig}, sCSI ${\mathcal H}_k=\{\gamma_{s,k},\kappa_{s,k},{\boldsymbol{\theta}}_{s,k},{\boldsymbol{\Sigma}}_{s,k}\}_{\forall s}$, which consists of slowly varying channel statistics, can be acquired and shared on a long time scale through pilot estimation, feedback, and ISLs \cite{you2020massive,li2021downlink,xiang2024massive,wu2025distributed,wang2026DP_JSAC}.
    Since only sCSI are required, the framework applies to both frequency-division duplexing (FDD) and time-division duplexing (TDD) systems.
    For SemComs, the multi-satellite massive MIMO system model introduces new spatial transmission degrees of freedom, while also raising new design considerations for the use of sCSI and the processing of spatial-domain signals and interference.

\vspace{-2mm}
\subsection{Received Signal Model of Multi-Satellite SemCom}
\vspace{-1mm}

In the considered system, distributed beamforming is employed to exploit the spatial multiplexing gain of the massive MIMO. For ease of exposition, we omit the indices of time-frequency resources and stack them column-wise to give an intuitive representation of the resources used by SemCom. 
With these expressions, the received signal at UT $k$ in \eqref{received signal eq1} is further expressed as
\begin{align}
    \textstyle{\bf Y}_{k} \!=\! \sum_{s\in\mathcal{S}_{k}}\!{\bf H}_{s,k}{\bf w}_{s,k} ({\bf z}_{s,k})^T + {\bf N}_{k}\in\mathbb{C}^{N_{\rm R}\times L},
    \label{received signal eq2}
\end{align}
where ${\bf w}_{s,k}\in\mathbb{C}^{N_{\rm T}\times 1}$ is the precoding vector from satellite $s$ to UT $k$, ${\bf z}_{s,k}\in\mathbb{C}^{L\times 1}$ is the semantic symbol vector transmitted from satellite $s$ to UT $k$, and $L$ denotes the per-satellite symbol length, i.e., the number of occupied time-frequency resources (e.g., subcarriers $\times$ symbols) at each serving satellite.
We assume that, under pre-compensation, the channel is approximately constant across the occupied resources of UT $k$.
In SemComs, $L$ is also related to the compression ratio (i.e., bandwidth ratio \cite{DeepJSCCMIMO2024}). For example, if $L$ complex symbols are used to transmit an RGB image of size $[3,H,W]$, the compression ratio is ${\rm CR}=(3HW)/L$ source samples per symbol.
Owing to the LoS-dominant nature of satellite channels, we use steering vectors to construct the distributed beamforming, i.e., ${\bf w}_{s,k}= \sqrt{P_{s,k}}\cdot{\bf v}^{*}_{s,k},\ \forall s,k$, where $P_{s,k}$ is the power-scaling factor chosen to satisfy the average transmit-power constraint $P_{\rm T}$.

This paper focuses on image-oriented SemCom to improve end-to-end reconstruction accuracy under constrained radio resources.
CT and NCT are two primary multi-satellite cooperative modes. For CT, multiple satellites transmit a common data stream, and the receiver demodulates the stream from the superimposed signal \cite{wang2026DP_JSAC,wu2025distributed}. 
In NCT, each satellite transmits an independent data stream \cite{xiang2024massive,cao2026DLMSCT}. Through the spatial-division capability of its multi-antenna array, the UT can support the transmission of multiple data streams from different satellites, but this process also introduces cross-stream interference.
These two modes correspond to two ways of exploiting the multi-satellite channel: the power gain from coherent combining and multiplexing gain from the increased number of spatially separated data streams at the receiver \cite{wang2026stf}. 
In Sections~\ref{framework ct sec},~\ref{framework nct sec}, and~\ref{sec:scalable_csi_module}, we develop the SemCom frameworks and network designs for MSCT, MSNCT, and MoCM, respectively.

\section{SemCom Framework for Multi-Satellite Coherent Transmission}
\label{framework ct sec}

\begin{figure*}[t]
    \centering
    \includegraphics[width=0.7\textwidth]{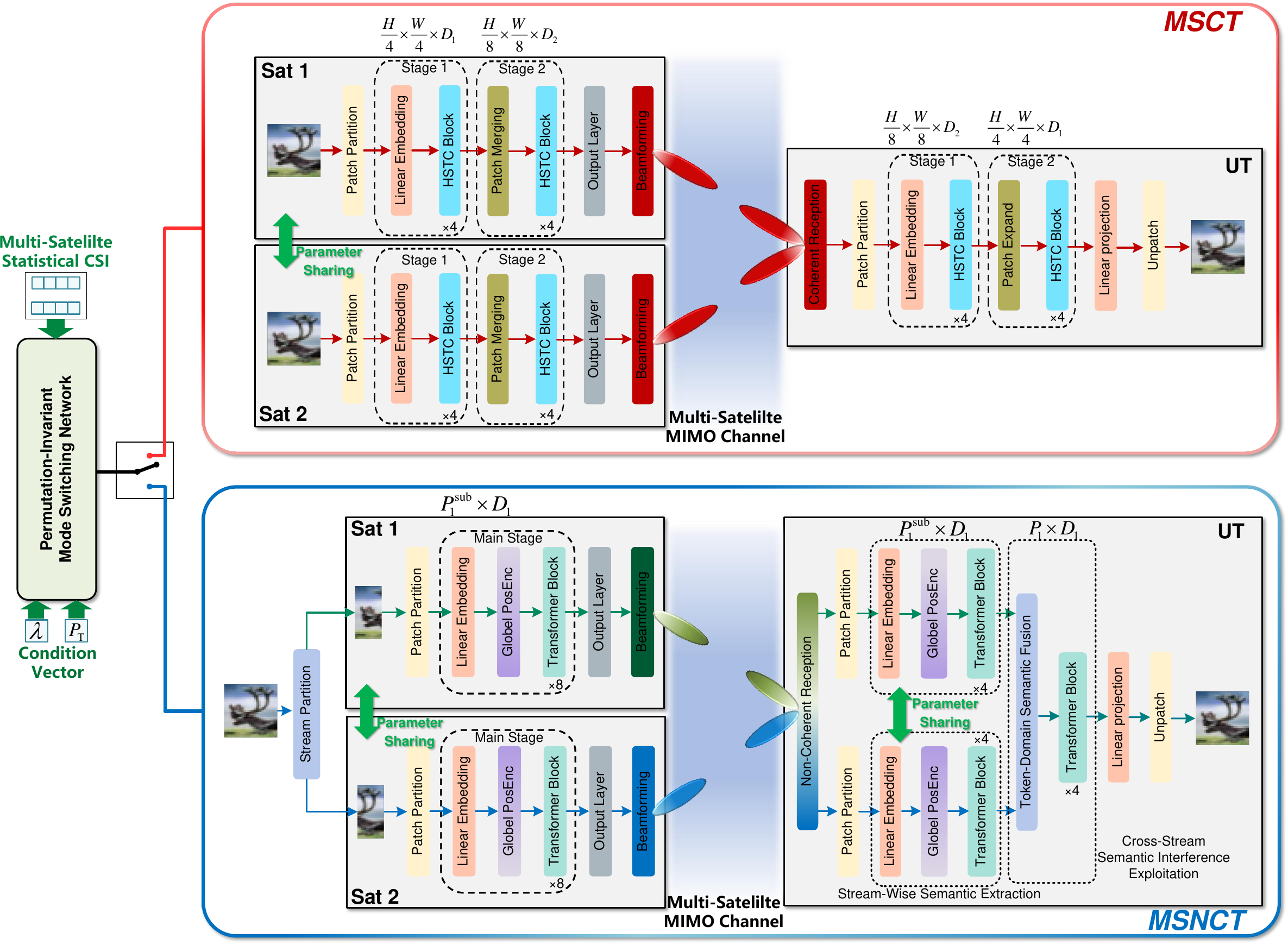}
    \vspace{-2mm}
    \caption{The proposed multi-satellite SemCom frameworks.}
    \label{fig:ms_semcom_framework}
    \vspace{-5mm}
\end{figure*}

Overall, the MSCT framework and its network design are shown in the upper part of Fig.~\ref{fig:ms_semcom_framework}.
\subsection{MSCT SemCom Architecture}

\subsubsection{Semantic Encoder}
In MSCT, all satellites convey the same semantic content with parameter-shared encoders \cite{wang2026DP_JSAC}. 
The encoders are deployed on board the satellites and thereby directly support the common case of on-board data generation in missions such as remote sensing.
Specifically, each satellite is equipped with an encoder instance using shared parameters, which preserves per-satellite adaptation capability, retains local decision flexibility, and naturally supports later extension to satellite-dependent adaptive encoding.
The semantic encoding can be written as
\vspace{-1mm}
\begin{align}
    {\bf z}_{k}=
    f^{\rm CT}_{\rm Enc}\big(\tI_k\big), \quad {\bf z}_{s,k}=
    {\bf z}_{k},
    \qquad \forall s\in\mathcal{S}_k,
    \label{eq:cen_sem_enc}
\end{align}
where $\tI_k\in\mathbb{R}^{C\times H\times W}$ denotes the source image for UT $k$. The encoder maps this image to the semantic symbol vector ${\bf z}_{s,k}\in\mathbb{C}^{L}$ with compression ratio $\mathrm{CR}=3HW/L$.

\subsubsection{Coherent Reception}
After the common semantic stream in \eqref{eq:cen_sem_enc} undergoes multi-satellite beamforming and channel propagation in \eqref{received signal eq2}, the receiver received signal matrix ${\bf Y}_k$.
Inspired by maximum-ratio combining, we exploit the geometric and sCSI to construct a low-complexity joint receive beam for combining the signals from the cooperating satellites, i.e.,
\begin{align}
    &\textstyle\tilde{\bf y}^T_{k}
    =
    \bar{\bf r}_{k}^{H}
    \left[
        \sum_{s\in\mathcal{S}_{k}}\!{\bf H}_{s,k}{\bf w}_{s,k} ({\bf z}_{k})^T
        + {\bf N}_{k}
    \right]
    \in\mathbb{C}^{1\times L}, \\
    &\textstyle\qquad 
    \bar{\bf r}_{k}
    =
    \sum_{s\in\mathcal{S}_{k}}\bar{\beta}_{s,k}{\bf u}_{s,k}, \ \ 
    \bar{\beta}_{s,k}
    =
    \frac{\beta_{s,k}}{\sum_{i\in\mathcal{S}_{k}}\beta_{i,k}},
\end{align}
where $ \beta_{s,k}
=
\sqrt{\frac{\gamma_{s,k}\kappa_{s,k}}{\kappa_{s,k}+1}}$, $\bar{\bf r}_{k}\in\mathbb{C}^{N_{\rm R}\times 1}$ is the joint receive beam for CT, and $\beta_{s,k}$ reflects the relative LoS strength of satellite $s$ inferred from the slowly varying sCSI.
Then, the receiver equalizes the received symbols and estimates the corresponding equivalent noise variance. Specifically,
\begin{align}
    &\textstyle\tilde{\bf y}^{\rm eq}_{k}
    =
    \zeta_k^{-1}\tilde{\bf y}_{k}
    \in
    \mathbb{C}^{L\times 1},\ \zeta_k
    =
    \bar{\bf r}_{k}^{H}\left(\sum_{s\in\mathcal{S}_{k}}\!{\bf H}_{s,k}{\bf w}_{s,k}\right), 
    \label{CT eq eq}\\
    &\sigma^{2}_{{\rm eq},k}
    =
    \mathbb{E}\{{\|\tilde{\bf y}_{k}-\zeta_k{\bf z}_k\|_2^2}\},
    \label{eq:ct_eq_feat}
\end{align}
where $\zeta_{k}\in\mathbb{C}$ and $\sigma^{2}_{{\rm eq},k}$ denote the effective demodulation channel and the noise variance that can be estimated from the downlink demodulation reference signal (DMRS), respectively. As the semantic decoder is inherently more robust to demodulation uncertainty than a conventional one, the estimation imperfection would further highlight its benefit.

\subsubsection{Semantic Decoder}
After coherent reception, $\tilde{\bf y}^{\rm eq}_{k}$ together with the low-dimensional equalization feature ${\boldsymbol{\delta}}_{k}\triangleq[\Re\{\zeta_{k}\},\,\Im\{\zeta_{k}\},\,\sigma^{2}_{{\rm eq},k}]^{T}\in\mathbb{R}^{3}$ is fed to the semantic decoder. The semantic reconstruction can be written as
\vspace{-1mm}
\begin{align}
    \hat{\tI}_k
    =
    g^{\rm CT}_{\rm Dec}\big(
        \tilde{\bf y}^{\rm eq}_{k},\,
        {\boldsymbol{\delta}}_{k}
    \big).
    \label{eq:ct_dec}
\end{align}

\begin{remark}
\vspace{-1mm}
MSCT Architecture Extension: The architecture can be further extended towards inter-satellite cooperation and the exploitation of sCSI. Specifically, the master satellite may dynamically orchestrate cooperative encoding by jointly leveraging the multi-satellite CSI and the source semantics~\cite{wang2026DP_JSAC}. 
Additionally, the cooperating satellites may also dynamically adjust their own encoding strategies according to their respective relative sCSI.
\end{remark}

\vspace{-2mm}
\subsection{Network Design for MSCT SemComs}
\vspace{-1mm}
To fully exploit the developed MSCT architecture, we design a matched Multi-Satellite Coherent Semantic Network (MSCSN). For the considered image semantic task, MSCSN adopts a symmetric encoder--decoder backbone stacked from hybrid Swin Transformer and CNN (HSTC) blocks, enabling hierarchical multi-resolution semantic interaction and receptive-field expansion. Compared with a Transformer-based backbone, the HSTC-based backbone inherits the favorable complexity-performance tradeoff and scalability of hierarchical Swin-Transformer designs \cite{swin-transformer,park2022how}. Relative to a pure Swin Transformer, it further incorporates a lightweight bottleneck CNN branch to complement window-based attention, better capture fine-grained local structures, and accelerate training convergence \cite{liu2023tcm}.

\subsubsection{Semantic Encoder}

Instead of the conventional separated source and channel coding scheme, we adopt an improved joint semantic encoder. Let the source semantic image be denoted by ${\bf X}_k\in\mathbb{R}^{3\times H\times W}$. The image is first partitioned into non-overlapping patches of size $p_1\times p_2$, which yields a total of $P_1=(H/p_1)(W/p_2)$ tokens. Denote by ${\bf x}_{k,p}\in\mathbb{R}^{3p_1p_2}$ the channel-wise vectorized representation of the $p$-th patch. The input processing can then be written as
\vspace{-1mm}
\begin{align}
    {\bf x}_{k,p}
    &=
    \mathrm{vec}\!\left(
        \mathrm{Patch}_{p}({\bf X}_k)
    \right)
    \in
    \mathbb{R}^{3p_1p_2},
    \ 
    p=1,\ldots,P_1, \\
    {\bf m}^{(0)}_{k,p}
    &=
    \mathrm{LN}\!\left(
        {\bf W}^{\rm CT}_{\rm in}{\bf x}_{k,p}
    \right)
    \in
    \mathbb{R}^{D_1},
\end{align}
where $\mathrm{LN}(\cdot)$ denotes layer normalization, and ${\bf W}^{\rm CT}_{\rm in}$ is the learnable input projection matrix of the CT semantic encoder.
Collecting all tokens yields the input sequence ${\bf M}^{(0)}_{k}\triangleq\{{\bf m}^{(0)}_{k,p}\}_{p=1}^{P_1}\in\mathbb{R}^{P_1\times D_1}$. No additional absolute positional embedding is introduced at this stage, because the adopted mixed backbone already preserves spatial structure explicitly.
The MSCSN encoder first operates on the full-resolution patch grid and then applies hierarchical patch merging to reach a lower-resolution but higher-dimensional latent grid. Let $P_2\triangleq P_1/4$ denote the number of second-stage latent tokens after one $2\times2$ patch-merging step. The resulting hidden representations can be summarized as
\vspace{-1mm}
\begin{align}
    {\bf M}^{(1)}_{k}
    &=
    \mathrm{MSCSN}_{\rm Enc,1}
    \bigl(
        {\bf M}^{(0)}_{k}
    \bigr)
    \in
    \mathbb{R}^{P_1\times D_1},
    \label{eq:ct_encoder_stage1_summary} \\
    {\bf Z}_{k}
    &=
    \mathrm{MSCSN}_{\rm Enc,2}
    \bigl(
        \mathrm{Merge}\bigl(
            {\bf M}^{(1)}_{k}
            \bigr)
        \bigr)
    \in
    \mathbb{R}^{P_2\times D_2},
    \label{eq:ct_encoder_stage2_summary} \\
    {\bf z}_{k}
    &=
    \mathrm{OutputLayer}
    \bigl(
        {\bf Z}_{k}
    \bigr)
    \in
    \mathbb{C}^{L},
    \label{eq:ct_encoder_hidden}
\end{align}
where $\mathrm{OutputLayer}(\cdot)$ denotes the output mapping from the deepest latent tokens to the transmitted symbol stream.

\begin{figure}[t]
    \centering
    \includegraphics[width=0.4\textwidth]{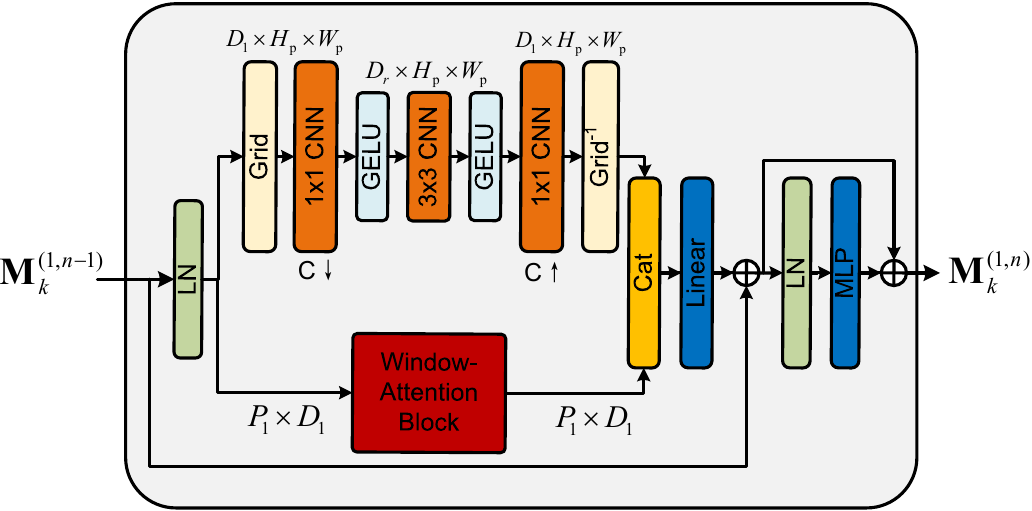}
    \vspace{-3mm}
    \caption{Structure of the HSTC block.}
    \label{fig:hstc_block}
    \vspace{-5mm}
\end{figure}

The first stage \eqref{eq:ct_encoder_stage1_summary} takes ${\bf M}^{(0)}_{k}\in\mathbb{R}^{P_1\times D_1}$ as its input and performs semantic interaction and feature refinement while preserving both the patch resolution $H_{\rm p}\times W_{\rm p}$ and the channel dimension $D_1$. Denoting by $N^{\rm CT}_{\rm Enc,1}$ the number of blocks in the first stage, the stage-1 processing can be written as
\begin{align}
    {\bf M}^{(1,n)}_{k}
    &\!\!=
    \mathrm{HSTC}^{(1,n)}
    \bigl(
        {\bf M}^{(1,n-1)}_{k}
    \bigr)
    \in
    \mathbb{R}^{P_1\times D_1},\  \forall n
\end{align}
with ${\bf M}^{(1,0)}_{k}\triangleq{\bf M}^{(0)}_{k}$, ${\bf M}^{(1)}_{k}\triangleq{\bf M}^{(1,N^{\rm CT}_{\rm Enc,1})}_{k}$, and $n=1,\ldots,N^{\rm CT}_{\rm Enc,1}$. This alternating pattern between regular-window attention and shifted-window attention is consistently adopted across the stacked HSTC blocks in both the first and second stages, thereby enlarging the effective receptive field while preserving the efficiency of localized computation \cite{swin-transformer,park2022how}. The internal structure of the $n$-th block in the first stage, illustrated in Fig.~\ref{fig:hstc_block}, follows as
\vspace{-1mm}
\begin{flalign}
    &\bar{\bf M}^{(1,n-1)}_{k}
    =
    \mathrm{LN}^{(1,n)}_{1}
    \bigl(
        {\bf M}^{(1,n-1)}_{k}
    \bigr)
    \in
    \mathbb{R}^{P_1\times D_1},
    \label{eq:ct_enc_stage1_ln1} &
    \\
    &{\bf A}^{(1,n)}_{k}
    =
    \mathrm{AttnBranch}^{(1,n)}
    \bigl(
        \bar{\bf M}^{(1,n-1)}_{k}
    \bigr)
    \in
    \mathbb{R}^{P_1\times D_1},
    \label{eq:ct_enc_stage1_attn} &
    \\
    &{\bf C}^{(1,n)}_{k}
    =
    \mathrm{ConvBranch}^{(1,n)}
    \bigl(
        \bar{\bf M}^{(1,n-1)}_{k}
    \bigr)
    \in
    \mathbb{R}^{P_1\times D_1},
    \label{eq:ct_enc_stage1_conv} &
    \\
    &{\bf G}^{(1,n)}_{k}
    =
    \mathrm{concat}\bigl(
        {\bf A}^{(1,n)}_{k},
        {\bf C}^{(1,n)}_{k}
    \bigr)
    \in
    \mathbb{R}^{P_1\times 2D_1},
    \label{eq:ct_enc_stage1_concat} &
    \\
    &\hat{\bf M}^{(1,n)}_{k}
    =
    {\bf M}^{(1,n-1)}_{k}
    +
    \mathrm{Linear}^{(1,n)}
    \bigl(
        {\bf G}^{(1,n)}_{k}
    \bigr),
    \label{eq:ct_enc_stage1_fuse} &
    \\
    &{\bf M}^{(1,n)}_{k}
    =
    \hat{\bf M}^{(1,n)}_{k}
    +
    \mathrm{MLP}^{(1,n)}
    \bigl(
        \mathrm{LN}^{(1,n)}_{2}
        \bigl(
            \hat{\bf M}^{(1,n)}_{k}
        \bigr)
    \bigr)
    \label{eq:ct_enc_stage1_block_out}, &
\end{flalign}
where $\mathrm{AttnBranch}(\cdot)$ and $\mathrm{ConvBranch}(\cdot)$ denote the Swin-Transformer attention branch and the lightweight bottleneck CNN branch, respectively. The two branch outputs are concatenated by $\mathrm{concat}(\cdot)$ and then fused through a linear projection to exploit the strengths of both branches, while $\mathrm{MLP}(\cdot)$ denotes a multi-layer perceptron (MLP).

In particular, the attention branch \eqref{eq:ct_enc_stage1_attn} adopts a Swin-Transformer module. Let the token-domain attention window size be $w_{\rm h}\times w_{\rm w}$, define $P_{\rm w}\triangleq w_{\rm h}w_{\rm w}$ as the number of tokens in each window, $N_{\rm h}$ as the number of attention heads, and $d_{\rm h}\triangleq D_1/N_{\rm h}$ as the per-head dimension. The normalized token sequence $\bar{\bf M}^{(1,n-1)}_{k}$ is first rearranged into a 2D token grid with spatial resolution $H_{\rm p}\times W_{\rm p}$. Then local self-attention is performed within each window.
For shifted-window blocks, a cyclic spatial shift is applied before window partition and is reversed after attention aggregation \cite{swin-transformer}. Let ${\bf X}^{(1,n)}_{k,r}\in\mathbb{R}^{P_{\rm w}\times D_1}$ denote the token matrix of the $r$-th window. Then the window-attention branch can be written as
\begin{alignat}{1}
    {\bf Q}^{(1,n)}_{k,r,h}
    &\!=\!
    {\bf X}^{(1,n)}_{k,r}
    {\bf W}^{Q,(1,n)}_{h}
    \!\in\!
    \mathbb{R}^{P_{\rm w}\times d_{\rm h}}, \\
    {\bf K}^{(1,n)}_{k,r,h}
    &\!=\!
    {\bf X}^{(1,n)}_{k,r}
    {\bf W}^{K,(1,n)}_{h}
    \!\in\!
    \mathbb{R}^{P_{\rm w}\times d_{\rm h}}, \\
    {\bf V}^{(1,n)}_{k,r,h}
    &\!=\!
    {\bf X}^{(1,n)}_{k,r}
    {\bf W}^{V,(1,n)}_{h}
    \!\in\!
    \mathbb{R}^{P_{\rm w}\times d_{\rm h}}, \\
    {\bf E}^{(1,n)}_{k,r,h}
    &\!=\!
    \frac{
        {\bf Q}^{(1,n)}_{k,r,h}
        \bigl(
            {\bf K}^{(1,n)}_{k,r,h}
        \bigr)^T
    }{\sqrt{d_{\rm h}}}
    \!+\!
    {\bf B}^{(1,n)}_{h}
    \!+\!
    {\bf \Omega}^{(1,n)}_{r}
    \!\!\in\!
    \mathbb{R}^{P_{\rm w}\times P_{\rm w}}, \\
    {\bf O}^{(1,n)}_{k,r,h}
    &\!=\!
    \mathrm{Softmax}
    \bigl(
        {\bf E}^{(1,n)}_{k,r,h}
    \bigr)
    {\bf V}^{(1,n)}_{k,r,h}
    \!\in\!
    \mathbb{R}^{P_{\rm w}\times d_{\rm h}}, \\
    \bar{\bf O}^{(1,n)}_{k,r}
    &\!=\!
    \mathrm{Proj}
    \bigl(
        \mathrm{Cat}_{h=1}^{N_{\rm h}}
        \!\bigl(
            {\bf O}^{(1,n)}_{k,r,h}
        \bigr)
    \bigr)
    \!\in\!
    \mathbb{R}^{P_{\rm w}\times D_1}, \\
    {\bf A}^{(1,n)}_{k}
    &\!=\!
    \Pi^{-1}_{\rm shift}
    \!\bigl(
        \mathrm{WinRev}
        \!\bigl(
            \bar{\bf O}^{(1,n)}_{k,r}
        \bigr)
    \bigr)
    \!\in\!
    \mathbb{R}^{P_1\times D_1}.
\end{alignat}
In the above expression, ${\bf W}^{Q,(1,n)}_{h}$, ${\bf W}^{K,(1,n)}_{h}$, and ${\bf W}^{V,(1,n)}_{h}$ belong to $\mathbb{R}^{D_1\times d_{\rm h}}$ and denote the learnable projections of the $h$-th head. Moreover, ${\bf B}^{(1,n)}_{h}\in\mathbb{R}^{P_{\rm w}\times P_{\rm w}}$ is the relative-position bias within each window, and ${\bf \Omega}^{(1,n)}_{r}\in\mathbb{R}^{P_{\rm w}\times P_{\rm w}}$ is the shifted-window mask. The operator $\mathrm{Cat}_{h=1}^{N_{\rm h}}(\cdot)$ concatenates all heads, $\mathrm{Proj}(\cdot)$ denotes the output linear projection, $\mathrm{WinRev}(\cdot)$ reassembles the window-wise token sequences into the 2D patch grid, and $\Pi^{-1}_{\rm shift}(\cdot)$ reduces to the identity mapping for non-shifted blocks. This branch therefore captures non-local semantic dependencies within each local region while preserving manageable computational complexity.

In parallel, the convolutional branch in \eqref{eq:ct_enc_stage1_conv} reshapes $\bar{\bf M}^{(1,n-1)}_{k}$ back to a feature map and processes it through a lightweight bottleneck CNN. Let $\bar{\bf F}^{(1,n-1)}_{k}\in\mathbb{R}^{D_1\times H_{\rm p}\times W_{\rm p}}$ denote the reshaped feature map and let $D_{r}\triangleq D_1/r_{\rm c}$ be the bottleneck width, where $r_{\rm c}$ is the channel-reduction ratio. Then the convolutional branch is given by
\begin{flalign}
    &\bar{\bf F}^{(1,n-1)}_{k}
    \!=\!
    \mathrm{Grid}
    \bigl(
        \bar{\bf M}^{(1,n-1)}_{k}
    \bigr)
    \!\in\!
    \mathbb{R}^{D_1\times H_{\rm p}\times W_{\rm p}}, \\
    &{\bf F}^{\rm pw,(1,n)}_{k}
    \!=\!
    \phi
    \bigl(
        {\bf W}^{\rm pw1,(1,n)}
        \circledast
        \bar{\bf F}^{(1,n-1)}_{k}
    \bigr)
    \!\in\!
    \mathbb{R}^{D_{r}\times H_{\rm p}\times W_{\rm p}}, \\
    &{\bf F}^{\rm dw,(1,n)}_{k}
    \!=\!
    \phi
    \bigl(
        {\bf W}^{\rm dw,(1,n)}
        \circledast
        {\bf F}^{\rm pw,(1,n)}_{k}
    \bigr)
    \!\in\!
    \mathbb{R}^{D_{r}\times H_{\rm p}\times W_{\rm p}}, \\
    &\tilde{\bf F}^{(1,n)}_{k}
    \!=\!
    {\bf W}^{\rm pw2,(1,n)}
    \circledast
    {\bf F}^{\rm dw,(1,n)}_{k}
    \!\in\!
    \mathbb{R}^{D_1\times H_{\rm p}\times W_{\rm p}}, \\
    &{\bf C}^{(1,n)}_{k}
    \!=\!
    \mathrm{Grid}^{-1}
    \bigl(
        \tilde{\bf F}^{(1,n)}_{k}
    \bigr)
    \!\in\!
    \mathbb{R}^{P_1\times D_1}.
\end{flalign}
where ${\bf W}^{\rm pw1,(1,n)}$ and ${\bf W}^{\rm pw2,(1,n)}$ are $1\times 1$ pointwise convolutions that first compress and then restore the channel dimension, ${\bf W}^{\rm dw,(1,n)}$ is a depthwise $3\times 3$ convolution, $\circledast$ denotes convolution, and $\phi(\cdot)$ denotes the GELU activation. 
Since CNN exhibits relatively large parameter count, the lightweight bottleneck design is developed to achieve a balance between representational capability and overfitting risk.

After the first-stage computation, the merge function $\mathrm{Merge}(\cdot)$ in \eqref{eq:ct_encoder_stage2_summary} further performs hierarchical downsampling by aggregating each non-overlapping $2\times 2$ neighborhood on the patch grid into a single latent token. Specifically, for latent-grid location $(i,j)$ and its associated latent token index $p$, the merge operation first concatenates the four neighboring tokens and then applies layer normalization followed by linear reduction. As a result, $\mathrm{Merge}(\cdot)$ reduces the latent resolution from $H_{\rm p}\times W_{\rm p}$ to $H_{\rm lat}\times W_{\rm lat}$ and produces $P_2=H_{\rm lat}W_{\rm lat}$ latent tokens. At the same time, it increases the channel dimension from $D_1$ to $D_2$. Let ${\bf M}^{(2)}_{k}\in\mathbb{R}^{P_2\times D_2}$ stacked from ${\bf m}^{(2)}_{k,p}$ denote the merged token sequence. The second stage \eqref{eq:ct_encoder_stage2_summary} contains $N^{\rm CT}_{\rm Enc,2}$ blocks and follows the same processing principle as the first stage, while its output is denoted by ${\bf Z}_{k}\in\mathbb{R}^{P_2\times D_2}$. More generally, the number of hierarchical stages can be scaled with the image resolution to support higher-resolution inputs.

The output layer then produces the final transmit symbols through the following computation:
\begin{align}
    {\bf U}_{k}
    =
    {\bf W}^{\rm CT}_{\rm sym}{\bf Z}_{k}
    \in
    \mathbb{R}^{P_2\times {2L/P_2}},\ 
    \tilde{\bf u}_{k}
    =
    \mathrm{RS}({\bf U}_{k})
    \in
    \mathbb{R}^{2L},\\
    \tilde{\bf z}_{k}
    =
    \tilde{\bf u}_{k[1:L]}
    +
    j\tilde{\bf u}_{k[L+1:2L]},\ 
    {\bf z}_{k}
    =
    \frac{\tilde{\bf z}_{k}}
    {\sqrt{
        \mathbb{E}\{\|{\tilde {\bf z}}_{k}\|_{\rm F}^2/L\}
    }}, 
\end{align}
where $\mathrm{RS}(\cdot)$ denotes the reshape operation.
Consequently, the transmit semantic signals in \eqref{eq:cen_sem_enc} are obtained as $\{{\bf z}_{s,k}\}_{s\in\mathcal{S}_k}$.

\subsubsection{Semantic Decoder}
The CT-oriented semantic decoder is designed symmetrically to the encoder.
After obtaining the equalized common stream in \eqref{CT eq eq}, the receiver first patches it into $P_2$ symbol tokens and then combines it with the equalization CSI feature and the noise statistics ${\boldsymbol{\delta}}_{k}$ defined before \eqref{eq:ct_dec} as inputs. The detailed expression is given by
\begin{align}
    {\bf d}^{(0)}_{k,p}
    &=
    \mathrm{Patchify}_{\rm sym}\!\left(
        \tilde{\bf y}^{\rm eq}_{k}
    \right)
    \in
    \mathbb{R}^{d^{\rm CT}_{\rm sym}},
    \qquad
    p=1,\ldots,P_2, \\
    {\bf q}^{(0)}_{k,p}
    &=
    {\bf W}^{\rm CT}_{\rm dec,in}
    \,
    \mathrm{concat}\bigl(
        {\bf d}^{(0)}_{k,p},
        \mathrm{LN}({\boldsymbol{\delta}}_{k})
    \bigr)
    \in
    \mathbb{R}^{D_2}.
    \label{eq:ct_decoder_input}
\end{align}
Stacking $\{{\bf q}^{(0)}_{k,p}\}_{p=1}^{P_2}$ along the token dimension yields ${\bf Q}^{(0)}_{k}\in\mathbb{R}^{P_2\times D_2}$. 
The decoder first refines these latent tokens and then expands them back to the full patch grid:
\begin{align}
    {\bf Q}^{\rm lat}_{k}
    &=
    \mathrm{MSCSN}_{\rm Dec,2}
    \!\bigl(
        {\bf Q}^{(0)}_{k}
    \bigr)
    \in
    \mathbb{R}^{P_2\times D_2}, \\
    {\bf Q}_{k}
    &=
    \mathrm{MSCSN}_{\rm Dec,1}
    \!\bigl(
        \mathrm{Expand}\!\left(
            {\bf Q}^{\rm lat}_{k}
        \right)
    \bigr)
    \in
    \mathbb{R}^{P_1\times D_1}.
    \label{eq:ct_decoder_hidden}
\end{align}
Here, $\mathrm{Expand}(\cdot)$ serves as the decoder-side counterpart of $\mathrm{Merge}(\cdot)$ by restoring the latent grid from resolution $H_{\rm lat}\times W_{\rm lat}$ to the full patch resolution $H_{\rm p}\times W_{\rm p}$, which is equivalently reflected by the token-domain transformation from $\mathbb{R}^{P_2\times D_2}$ to $\mathbb{R}^{P_1\times D_1}$.
Finally, each refined token is projected back to a patch-level RGB vector and rearranged to form the reconstructed image,
\begin{align}
    \hat{\bf i}_{k,p}
    &=
    {\bf W}^{\rm CT}_{\rm rec}\,
    {\bf Q}_{k[p,:]}
    \in \mathbb{R}^{3p_1p_2},
    \qquad p=1,\ldots,P_1, \\
    \hat{\tI}_{k}
    &=
    \mathrm{Unpatch}
    \bigl(
        \{\hat{\bf i}_{k,p}\}_{p=1}^{P_1}
    \bigr)
    \in \mathbb{R}^{3\times H\times W},
    \label{eq:ct_decoder_output}
\end{align}
where $\mathrm{Unpatch}(\cdot)$ is the inverse of $\mathrm{Patchify}(\cdot)$ by rearranging patches according to the original image grid.

\subsubsection{Complexity Analysis}\label{sec:ct_complexity}

We characterize the complexity from three aspects. Regarding user-data acquisition, under the distributed realization adopted in \eqref{eq:cen_sem_enc}, every serving satellite has to access the entire user information $\tI_k$. For the inference-stage computational complexity, the dominant cost is contributed by the linear projections, MLPs, and local window self-attention operations in the networks. Accordingly, letting $P_{\rm w}\triangleq w_{\rm h}w_{\rm w}$ denote the number of tokens within each attention window, the complexity incurred at each satellite scales as $\mathcal{O}\big(N^{\rm CT}_{\rm Enc,1}(P_1D_1^2+P_1P_{\rm w}D_1)+N^{\rm CT}_{\rm Enc,2}(P_2D_2^2+P_2P_{\rm w}D_2)\big)$. Likewise, if $N^{\rm CT}_{\rm Dec,1}$ and $N^{\rm CT}_{\rm Dec,2}$ denote the numbers of decoder blocks in the two stages, the receiver-side complexity is of order $\mathcal{O}\big(N^{\rm CT}_{\rm Dec,2}(P_2D_2^2+P_2P_{\rm w}D_2)+N^{\rm CT}_{\rm Dec,1}(P_1D_1^2+P_1P_{\rm w}D_1)\big)$. For the network parameter size, the dominant contribution likewise comes from the learnable linear and pointwise transformations. Therefore, the transmitter and receiver parameter counts are of orders $\mathcal{O}(N^{\rm CT}_{\rm Enc,1}D_1^2+N^{\rm CT}_{\rm Enc,2}D_2^2)$ and $\mathcal{O}(N^{\rm CT}_{\rm Dec,1}D_1^2+N^{\rm CT}_{\rm Dec,2}D_2^2)$, respectively.
Owing to its lightweight design, the complexity of the bottleneck CNN is negligible and therefore ignored.

\section{Semantic Communication Framework for Multi-Satellite Non-Coherent Transmission}
\label{framework nct sec}
Unlike MSCT, MSNCT involves different semantic streams transmitted by different satellites and mutually interfering with each other. Therefore, the key design issue is how to preserve the distinctiveness of different streams while fully exploiting their complementary semantic relationship to handle semantic-domain interference.
The overall MSNCT architecture and its network design are summarized in the lower part of \figref{fig:ms_semcom_framework}.

\vspace{-2mm}
\subsection{MSNCT SemCom Architecture}
\label{NCT architecture sec}
\vspace{-1mm}

\subsubsection{Semantic Encoder}
The source content is first distributed by the master satellite and then assigned to other satellites. This distribution can be performed across an image set or within an individual image, while the implementation below instantiates it as image partitioning. Each satellite subsequently processes its assigned content through a parameter-shared encoder $f^{\rm NCT}_{\rm Enc}(\cdot)$ to generate the transmit symbols. The overall encoding can be expressed as
\vspace{-3mm}
\begin{align}
    \{\tI_{s,k}\}_{s\in\mathcal{S}_k}
    =
    f^{\rm NCT}_{\rm Distribute}\big(\tI_k\big)
    \label{eq:nct_sem_distribute},\\
    {\bf z}_{s,k}
    =
    f^{\rm NCT}_{\rm Enc}\big(\tI_{s,k}\big),
    \quad
    \forall s\in\mathcal{S}_k.
    \label{eq:nct_sem_enc}
\end{align}

\subsubsection{Non-Coherent Reception}
After the per-satellite data streams traverse the multi-satellite transmitter and the wireless channel \eqref{received signal eq2}, the UT receives the signal matrix ${\bf Y}_k$.
The distinct linear receiver for each satellite is used to extract the corresponding signal to be demodulated, i.e.,
\begin{equation}
    \textstyle{\tilde {\bf y}}^T_{s,k} = {\bf r}^H_{s,k} \left[\sum_{i\in\mathcal{S}_{k}}\!{\bf H}_{i,k}{\bf w}_{i,k} ({\bf z}_{i,k})^T + {\bf N}_{k}\right]\in\mathbb{C}^{1\times L}.
\end{equation}
where ${\bf r}_{s,k}\in\mathbb{C}^{N_{\rm R}\times 1}$ is the receive combiner at UT $k$ for the signal from satellite $s$. 
Similar to transmit beamforming, we set the receive beamformer as the receive steering vector, i.e., ${\bf r}_{s,k} = {\bf u}_{s,k}, \ \forall s,k$, where ${\bf u}_{s,k}$ is determined by the geometric AoA of satellite $s$ at UT $k$ contained in the sCSI.
We further eliminate the residual amplitude and phase distortion on each detected stream by equalization, i.e.,
\begin{align}
    &\tilde{\bf y}^{\rm eq}_{s,k}
    =
    \zeta^{-1}_{s,k}\tilde{\bf y}_{s,k}
    \in
    \mathbb{C}^{L\times 1}, \ \zeta_{s,k}=
    {\bf r}^{H}_{s,k}{\bf H}_{s,k}{\bf w}_{s,k}, \\
    &\sigma^{2}_{{\rm eq},s,k}
    =
    \mathbb{E}\bigl\{
        \|\tilde{\bf y}^{\rm eq}_{s,k}-{\bf z}_{s,k}\|_2^2
    \bigr\},
    \qquad
    \forall s\in\mathcal{S}_k,
    \label{eq:nct_eq_feat}
\end{align}
where $\zeta_{s,k}\in\mathbb{C}$ denotes the effective demodulation channel of stream $s$, and $\sigma^{2}_{{\rm eq},s,k}$ denotes the corresponding equivalent noise variance after equalization. Estimating these block-level features in deployment does not require knowledge of the transmitted semantic symbols. Different from the MSCT, this equivalent-noise term absorbs not only the noise but also the residual leakage from other semantic streams.

\subsubsection{Semantic Decoder}
Multiple satellites transmit distinct data streams, which mutually interfere. In contrast to the conventional scheme that demodulates each stream independently, we propose a two-stage demodulation framework, in which the $S_k$ data streams first undergo parameter-shared stream-wise semantic extraction individually, after which the semantics are jointly processed to utilize the cross-stream semantic coupling for interference exploitation. The semantic decoder is expressed as
\begin{align}
    &{\tilde \tQ}_{s,k}
    =
    g^{\rm NCT}_{\rm Dec,SW}\big(
        \tilde{\bf y}^{\rm eq}_{s,k},
        {\boldsymbol{\delta}}_{s,k},{\bf Y}_k
    \big),
    \quad
    \forall s\in\mathcal{S}_k,
    \label{eq:nct_dec_early}\\
    &\tQ_k
    =
    f^{\rm NCT}_{\rm merge}\big(
        \{{\tilde \tQ}_{s,k}\}_{s\in\mathcal{S}_k}
    \big),\ \ 
    \hat{\tI}_k
    =
    g^{\rm NCT}_{\rm Dec,CS}\big(
        \tQ_k
    \big),
    \label{eq:nct_dec_late}
\end{align}
where both the equalization-related feature ${\boldsymbol{\delta}}_{s,k}
\triangleq
\big[
    \Re\{\zeta_{s,k}\},
    \Im\{\zeta_{s,k}\},
    \sigma^{2}_{{\rm eq},s,k}
\big]^{T}$ and the raw received signal ${\bf Y}_k$ are fed in to prevent potentially useful information from being suppressed by the receive beamforming.
Rather than directly applying joint feature detection to the mixed received signal, we adopt the two-stage receiver in \eqref{eq:nct_dec_early}--\eqref{eq:nct_dec_late} for stability and scalability. The first stage recovers each satellite's stream separately, preserving symmetry with the transmitter. Moreover, as the complexity of strong backbones often grows superlinearly with the token count, exemplified by the quadratic floating-point operation (FLOP) scaling of Transformer \cite{vit}, processing shorter per-stream sequences in the first stage and judiciously activating the second stage yields a favorable performance-complexity tradeoff.

\begin{remark}
    MSNCT Architecture Extension: The architecture can be naturally extended to support dynamic cooperative stream partition and encoding by jointly exploiting satellite-specific CSI and the source semantics. In \eqref{eq:nct_sem_distribute} and \eqref{eq:nct_sem_enc}, the master satellite may incorporate sCSI into the partition rule so as to adaptively determine how the semantic content is split and assigned across the cooperating satellites.
\end{remark}

\vspace{-3mm}
\subsection{Network Design for MSNCT SemCom}
\vspace{-1mm}
The crux of the network of MSNCT lies in how to handle cross-stream interference induced by link-level interference, while accounting for the semantic correlation among streams partitioned from the same source image.
Compared with HSTC, we adopt the Transformer block of comparable scale as the basic module \cite{vit,DeepJSCCMIMO2024}, because the global attention of the latter affords markedly stronger semantic interaction capability, thereby offering a notable performance advantage.

\subsubsection{Stream Partition and Semantic Encoder}

Without loss of generality, we adopt a direct width-wise equal-partition scheme as $f^{\rm NCT}_{\rm Distribute}(\cdot)$. More sophisticated partitioning rules, such as Discrete Cosine Transform- or wavelet-domain splitting used in image compression, can also be incorporated into $f^{\rm NCT}_{\rm Distribute}(\cdot)$ to allocate different transform components across satellites. Specifically, the partition module $f^{\rm NCT}_{\rm Distribute}$ splits the full image along the width dimension into $S_k$ equal sub-images. For any $s\in\mathcal{S}_k$, we have
\vspace{-1mm}
\begin{align}
    \tI_{s,k}
    &=
    \mathrm{Crop}_{s}(\tI_k)
    \in
    \mathbb{R}^{3\times H\times W_{\rm sub}},
    \ s\in\mathcal{S}_k,
    \label{eq:nct_half_split}
\end{align}
where $W_{\rm sub}\triangleq W/S_k$, $S_k\triangleq|\mathcal{S}_k|$, and $\mathrm{Crop}_{s}(\cdot)$ extracts the $s$-th width-wise sub-image of $\tI_k$. The resulting set $\{\tI_{s,k}\}_{s\in\mathcal{S}_k}$ provides a natural semantic partition for MSNCT. For illustration, Fig.~\ref{fig:ms_semcom_framework} depicts the case with $S_k=2$.

In the encoder, each cooperating satellite employs a parameter-shared Transformer-based backbone and performs semantic encoding solely on its own allocated sub-image. Since the tokens of different streams possess an inherent geometric ordering prior to partitioning, we inject global positional embeddings into each sub-image encoder rather than identical ones to preserve the global geometric prior consistent with the full image, thereby laying the foundation for subsequent joint cross-stream processing. Let $P^{\rm sub}\triangleq(H/p_1)(W_{\rm sub}/p_2)=P_1/S_k$ denote the number of patch tokens in each semantic stream, and let ${\bf x}_{s,k,p}\in\mathbb{R}^{3p_1p_2}$ denote the $p$-th patch of sub-image $\tI_{s,k}$. Let further $\pi_s(p)\in\{1,\ldots,P_1\}$ denote the row-major index of that patch within the full-image grid. The input embedding stage then reads
\vspace{-1mm}
\begin{align}
    {\bf m}^{(0)}_{s,k,p}
    &=
    \mathrm{LN}\bigl(
        {\bf W}^{\rm NCT}_{\rm in}
        {\bf x}_{s,k,p}
    \bigr)
    +
    {\bf e}_{\pi_s(p)}
    \in
    \mathbb{R}^{D_1},
\end{align}
where $p=1,\ldots,P^{\rm sub}$ and $\{{\bf e}_{j}\}_{j=1}^{P_1}\subset\mathbb{R}^{D_1}$ is the learnable full-image positional embedding table shared by all streams. Collecting all tokens of stream $s$ yields ${\bf M}^{(0)}_{s,k}\triangleq\{{\bf m}^{(0)}_{s,k,p}\}_{p=1}^{P^{\rm sub}}\in\mathbb{R}^{P^{\rm sub}\times D_1}$. The shared Transformer-based encoder then performs semantic feature refinement by stacking $N^{\rm NCT}_{\rm Enc}$ Transformer blocks:
\vspace{-1mm}
\begin{align}
    {\bf M}^{(n)}_{s,k}
    &=
    \mathrm{TF}^{(n)}_{\rm Enc}
    \bigl(
        {\bf M}^{(n-1)}_{s,k}
    \bigr)
    \in
    \mathbb{R}^{P^{\rm sub}\times D_1},
    \ \forall n,
    \label{eq:nct_enc_blocks}
\end{align}
\vspace{-1mm}
with ${\bf Z}_{s,k}\triangleq{\bf M}^{(N^{\rm NCT}_{\rm Enc})}_{s,k}$ and $n=1,\ldots,N^{\rm NCT}_{\rm Enc}$.
Each block $\mathrm{TF}^{(n)}_{\rm Enc}$ uses a Transformer encoder structure with an MHSA sublayer and an MLP sublayer, each equipped with residual connection and post-layer normalization \cite{DeepJSCCMIMO2024}.

Similar to the CT output layer, each patch token is independently projected and then power-normalized to form the transmit symbol stream as
\vspace{-1mm}
\begin{align}
    &{\bf U}_{s,k}
    =
    {\bf W}^{\rm NCT}_{\rm sym}
    {\bf Z}_{s,k}
    \in
    \mathbb{R}^{P^{\rm sub}\times \frac{2L}{P^{\rm sub}}},\ 
    \tilde{\bf u}_{s,k}
    =
    \mathrm{RS}({\bf U}_{s,k})
    , \\
    &\tilde{\bf z}_{s,k}
    \!=\!
    \tilde{\bf u}_{s,k[1:L]}
    \!+\!
    j\tilde{\bf u}_{s,k[L+1:2L]},\ 
    {\bf z}_{s,k}
    \!=\!
    \frac{\tilde{\bf z}_{s,k}}
    {\sqrt{
        \mathbb{E}\{\|{\tilde {\bf z}}_{s,k}\|_{\rm F}^2/L\}
    }}.
\end{align}
The key difference is that each satellite ultimately produces a symbol stream carrying different semantics.

\subsubsection{Stream-Wise Semantic Extraction}

At the receiver, we employ a Transformer network including $N^{\rm NCT}_{\rm Dec,SW}$ layers to build $f^{\rm NCT}_{\rm Dec, SW}$. Beyond the equalized demodulated symbols of its own stream, each per-stream token further incorporates two types of auxiliary information: the raw received-signal context ${\bf Y}_k$ shared across all streams, and the equivalent channel of the current stream. 
\vspace{-1mm}
\begin{align}
    {\bf d}^{(0)}_{s,k,p}
    &=
    \mathrm{Patchify}_{\rm sym}\!\bigl(
        \tilde{\bf y}^{\rm eq}_{s,k}
    \bigr)
    \in
    \mathbb{R}^{\frac{2L}{P^{\rm sub}}},\\
    {\bf r}^{(0)}_{k,p}
    &=
    \gamma
    \,\mathrm{LN}\!\left(
        \mathrm{Patchify}_{\rm rx}\!\left(
            {\bf Y}_{k}
        \right)
    \right)
    \in
    \mathbb{R}^{\frac{2N_{\rm R}L}{P^{\rm sub}}},
    \\
    {\boldsymbol{\xi}}^{(0)}_{s,k,p}
    &=
    \mathrm{concat}\bigl(
        {\bf d}^{(0)}_{s,k,p},
        {\bf r}^{(0)}_{k,p},
        \mathrm{LN}\left(
        {\boldsymbol{\delta}}_{s,k}\right)
    \bigr)
    \in
    \mathbb{R}^{d^{\rm NCT}_{\rm tok}}, \\
    {\bf q}^{(0)}_{s,k,p}
    &=
    {\rm MLP}(
    {\boldsymbol{\xi}}^{(0)}_{s,k,p}
    )
    +
    {\bf e}_{\pi_s(p)}
    \in
    \mathbb{R}^{D_1},
    \label{eq:nct_decoder_input}
\end{align}
where ${\rm MLP}$ here refers to the input mapper constructed by Siam-style MLP \cite{DeepJSCCMIMO2024}.
$d^{\rm NCT}_{\rm tok}\triangleq d^{\rm NCT}_{\rm sym}+d^{\rm NCT}_{\rm rx}+3$. $\gamma$ is a zero-initialized learnable scalar gating the raw received-signal branch, which stabilizes the early training and allows its contribution to grow as learning progresses. Collecting all stream-wise tokens yields
\vspace{-2mm}
\begin{align}
    {\bf Q}^{(0)}_{s,k}
    &\triangleq
    \{{\bf q}^{(0)}_{s,k,p}\}_{p=1}^{P^{\rm sub}}
    \in
    \mathbb{R}^{P^{\rm sub}\times D_1},\ s\in\mathcal{S}_k.
\end{align}
A parameter-shared Transformer-based backbone is then applied to all semantic streams by stacking $N^{\rm NCT}_{\rm Dec,SW}$ blocks:
\vspace{-2mm}
\begin{align}
    {\bf Q}^{(n)}_{s,k}
    &=
    \mathrm{TF}^{(n)}_{\rm Dec,SW}
    \!\bigl(
        {\bf Q}^{(n-1)}_{s,k}
    \bigr)
    \in
    \mathbb{R}^{P^{\rm sub}\times D_1},\ \forall n.
    \label{eq:nct_decoder_hidden_early}
\end{align}
with $n=1,\ldots,N^{\rm NCT}_{\rm Dec,SW}$ and $s\in\mathcal{S}_k$.

\subsubsection{Token-Domain Semantic Fusion}

After the initial per-stream semantic extraction, we carry out the semantic fusion $f^{\rm NCT}_{\rm merge}(\cdot)$ in the token domain by concatenating on the patch grid, rather than via MLP-based fusion in the latent space. This design preserves the information of each semantic stream while leveraging the semantic interaction capability of the global attention of Transformer blocks in the token domain, which is the key to exploiting cross-stream interference. 
Concretely, let $H_{\rm p}\triangleq H/p_1$ and $W_{\rm p,sub}\triangleq W_{\rm sub}/p_2=W_{\rm p}/S_k$, and reshape each stream's latent token sequence back onto its sub-image patch grid:
\vspace{-1mm}
\begin{align}
    {\tilde {\bf Z}}_{s,k}
    &\textstyle=
    \mathrm{RS}(
        \tilde{\bf Q}_{s,k}
    )
    \in
    \mathbb{R}^{H_{\rm p}\times W_{\rm p,sub}\times D_1},
    \ s\in\mathcal{S}_k,
\end{align}
where $\tilde{\bf Q}_{s,k}\triangleq{\bf Q}^{(N^{\rm NCT}_{\rm Dec,SW})}_{s,k}$.
These tokens are restored to their positions in the token domain according to the geometric relationships, followed by flattening into a patch-token sequence:
\vspace{-5mm}
\begin{align}
    {\tilde {\bf Z}}_{k}
    &\textstyle=
    \mathrm{concat}_{\rm width}
    (
        \{{\tilde {\bf Z}}_{s,k}\}_{s\in\mathcal{S}_k}
    )
    \in
    \mathbb{R}^{H_{\rm p}\times W_{\rm p}\times D_1}, \\
    {\tilde {\bf Q}}_{k}
    &\textstyle=
    \mathrm{flatten}(
        {\tilde {\bf Z}}_{k})
    \in
    \mathbb{R}^{P_1\times D_1}.
    \label{eq:nct_tokcat_merge}
\end{align}
Here, $\mathrm{concat}_{\rm width}(\cdot)$ assembles the stream-wise grids along the width dimension in the same order used by \eqref{eq:nct_half_split}. Although the concrete implementation in this paper reduces to the width-wise concatenation, the general architecture in Section~\ref{NCT architecture sec} still allows $f^{\rm NCT}_{\rm merge}$ to be extended to a CSI-aware fusion function under other stream-partition strategies.

\subsubsection{Cross-Stream Semantic Interference Exploitation}
$g^{\rm NCT}_{\rm Dec, CS}(\cdot)$ operates on the full-image token grid with a stack of $N^{\rm NCT}_{\rm Dec, CS}$ Transformer blocks to perform joint semantic reconstruction across streams:
\begin{align}
    {\bf Q}^{(n)}_{k}
    &=
    \mathrm{TF}^{(n)}_{\rm Dec,CS}
    \!\bigl(
        {\bf Q}^{(n-1)}_{k}
    \bigr)
    \in
    \mathbb{R}^{P_1\times D_1},
    \ \forall n,\\
    \hat{\bf i}_{k,p}
    &=
    {\bf W}^{\rm NCT}_{\rm rec}\,
    {\bf Q}^{(N^{\rm NCT}_{\rm Dec,CS})}_{k[p,:]}
    \in
    \mathbb{R}^{3p_1p_2},
    \  p=1,\ldots,P_1, \\
    \hat{\tI}_{k}
    &=
    \mathrm{Unpatch}
    \bigl(
        \{\hat{\bf i}_{k,p}\}_{p=1}^{P_1}
    \bigr)
    \in
    \mathbb{R}^{3\times H\times W},
    \label{eq:nct_decoder_output}
\end{align}
where ${\bf Q}^{(0)}_{k} \triangleq {\tilde{\bf Q}}_k$ and $\mathrm{Unpatch}(\cdot)$ denotes the inverse of $\mathrm{Patchify}(\cdot)$, which rearranges the recovered patches according to the original image grid. The self-attention mechanism within the Transformer at this stage enables tokens of one stream to interact with those from any other stream, thereby converting destructive cross-stream semantic interference into constructive semantic enhancement, a process that parallels interference exploitation \cite{wang2022weighted} and benefits from token-domain processing for reconstruction.

\subsubsection{Complexity Analysis}\label{sec:nct_complexity}

With respect to user-data acquisition, each satellite under MSNCT only needs to acquire a portion of the user source, whose average size is $1/S_k$ of that required by a single CT satellite. Regarding the inference-stage computational complexity, because each transmitter encodes only its allocated stream with $P^{\rm sub}=P_1/S_k$ tokens, the per-satellite transmit-side complexity is $\mathcal{O}\bigl(N^{\rm NCT}_{\rm Enc}(P^{\rm sub} D_1^2 + (P^{\rm sub})^2 D_1)\bigr)$, which scales approximately as $1/S_k$ of the CT transmit-side complexity in the linear projection-dominated regime. At the receiver, $g^{\rm NCT}_{\rm Dec, SW}(\cdot)$ is applied in parallel to the $S_k$ streams, each processing $P^{\rm sub}$ tokens, whereas $g^{\rm NCT}_{\rm Dec,CS}(\cdot)$ jointly processes all $P_1$ tokens. The overall receiver-side complexity is therefore $\mathcal{O}\bigl(S_k N^{\rm NCT}_{\rm Dec,SW}(P^{\rm sub} D_1^2 + (P^{\rm sub})^2 D_1) + N^{\rm NCT}_{\rm Dec, CS}(P_1 D_1^2 + P_1^2 D_1)\bigr)$. In terms of network parameter size, the transmit-side and receive-side parameter sizes are $\mathcal{O}(N^{\rm NCT}_{\rm Enc} D_1^2)$ and $\mathcal{O}\bigl((N^{\rm NCT}_{\rm Dec,SW} + N^{\rm NCT}_{\rm Dec,CS}) D_1^2\bigr)$, respectively, and neither grows with $S_k$.

\vspace{-2mm}
\section{Mixture of Cooperative Modes (MoCM) for Multi-Satellite SemComs}
\label{sec:scalable_csi_module}
\vspace{-1mm}

MSCT and MSNCT offer complementary gains in link-budget enhancement and stream multiplexing, while also differing in complexity, with MSNCT requiring substantially less user-data acquisition and satellite-side computation. 
To balance these tradeoffs, this section proposes, for the first time, a mixture-of-experts (MoE)-inspired MoCM framework for multi-satellite SemComs \cite{du2022glam}, which treats CT and NCT as mode-level experts and adaptively selects the more suitable mode according to channel conditions and the performance-complexity tradeoff.
The overall framework is illustrated in \figref{fig:ms_semcom_framework}.
This paper focuses on the tradeoff between a task-level semantic performance metric and satellite-side complexity, where the metric can be chosen according to the target application. In view of the substantially lower satellite-side complexity of MSNCT relative to MSCT, we define a selection label that gives priority to MSNCT whenever it can satisfy a prescribed performance requirement in comparison with MSCT, i.e., 
\begin{align}
    m_k^{\star}
    =
    \begin{cases}
        1, & \mathcal{Q}^{\rm NCT}_{k} \ge \lambda\,\mathcal{Q}^{\rm CT}_{k}, \\
        0, & \text{otherwise},
    \end{cases}
    \label{eq:selector_oracle}
\end{align}
where $\mathcal{Q}^{\rm NCT}_{k}$ and $\mathcal{Q}^{\rm CT}_{k}$ denote the selected task-level performance metric achieved by the NCT and CT modes, respectively, and $\lambda\in[0,1]$ is the prescribed tradeoff factor. A larger $\lambda$ places more emphasis on semantic performance, whereas a smaller $\lambda$ gives higher priority to complexity reduction. In our implementation, $\mathcal{Q}$ is instantiated as the reconstruction peak signal-to-noise ratios (PSNRs), but other task-oriented metrics can be used by regenerating the labels. The metric values in \eqref{eq:selector_oracle} are not assumed to be available before mode selection. They are used only to generate offline supervision labels, while the online mode-switching module predicts $m_k^{\star}$ from the available sCSI representation developed below.

\vspace{-2mm}
\subsection{Scalable CSI Representation}
\vspace{-1mm}

Due to the high mobility of satellites, accurate instantaneous CSI is difficult to obtain in practice. We therefore use the slowly varying sCSI  ${\mathcal H}_k=\{\gamma_{s,k},\kappa_{s,k},{\boldsymbol{\theta}}_{s,k},{\boldsymbol{\Sigma}}_{s,k}\}_{\forall s}$ to construct the input of the mode-switching module. Although statistical CSI is already much lower-dimensional than instantaneous CSI, the receive-side covariance matrix ${\boldsymbol{\Sigma}}_{s,k}\in\mathbb{C}^{N_{\rm R}\times N_{\rm R}}$ remains high-dimensional. When the number of receive antennas becomes large, directly flattening this matrix into the network input would lead to a quadratic increase in both parameter count and computational complexity. To address this issue, we process it as follows.
\begin{align}
    q^{\Sigma}_{s,k}
    &\triangleq
    {\bf d}_{s,k}^{H}
    {\boldsymbol{\Sigma}}_{s,k}
    {\bf d}_{s,k}
    \in
    \mathbb{R}_{+},
\end{align}
where ${\bf d}_{s,k}\in\mathbb{C}^{N_{\rm R}}$ is the normalized receive steering vector at the receiver, and $q^{\Sigma}_{s,k}\in\mathbb{R}_{+}$ denotes the covariance power projected onto the receive-beam direction. This quantity characterizes the level of NLoS interference aligned with the main beam after beamforming. Through this operation, the covariance-related feature dimension is reduced from the $N_{\rm R}^{2}$ dimensions of direct vectorization to a single scalar.
\vspace{-2mm}
\begin{align}
    {\boldsymbol{\psi}}^{\rm sel}_{s,k}
    &\triangleq
    \left[
        \gamma_{s,k},
        {\bar \kappa}_{s,k},
        \theta^{\rm t}_{s,k},
        \phi^{\rm t}_{s,k},
        \theta^{\rm r}_{s,k},
        \phi^{\rm r}_{s,k},
        {\bar q}^{\Sigma}_{s,k}
    \right]^{T}
    \in \mathbb{R}^{7},
    \label{eq:selector_scsi_vector}
\end{align}
where ${\bar \kappa}_{s,k}=10\log_{10}\kappa_{s,k}$, and ${\bar q}^{\Sigma}_{s,k}=\log \textstyle(q^{\Sigma}_{s,k})$.

\vspace{-2mm}
\subsection{Permutation-Invariant Mode Switching Network}
\vspace{-1mm}

\begin{figure}[t]
    \centering
    \includegraphics[width=0.3\textwidth]{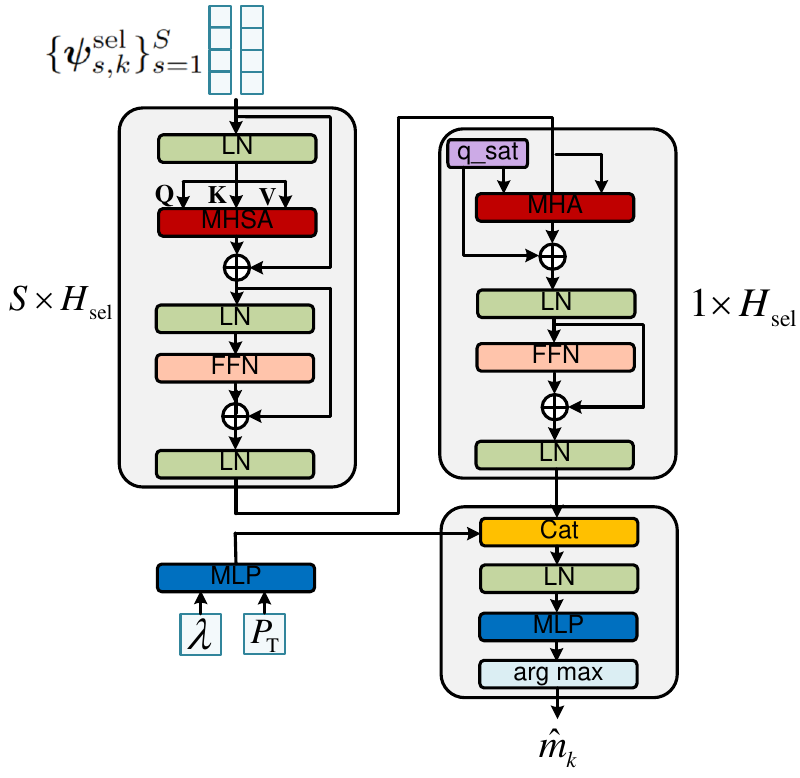}
    \vspace{-5mm}
    \caption{Permutation-invariant mode-switching network.}
    \label{fig:selector_fig}
    \vspace{-5mm}
\end{figure}

The overall architecture of the proposed permutation-invariant mode-switching network is illustrated in Fig.~\ref{fig:selector_fig}.
The per-satellite features $\{{\boldsymbol{\psi}}^{\rm sel}_{s,k}\}_{s=1}^{S_k}$ are treated as an unordered set rather than an ordered tuple, because the CT/NCT mode decision should not depend on the order of the satellites. This means that the mapping from multi-satellite features to the mode decision must be permutation invariant \cite{11049893}, i.e., $f_{\rm sel}(
    \{{\boldsymbol{\psi}}^{\rm sel}_{s,k}\}_{s=1}^{S_k})
=
f_{\rm sel}(
    \{{\boldsymbol{\psi}}^{\rm sel}_{\pi(s),k}\}_{s=1}^{S_k}),$
where $\pi$ denotes an arbitrary permutation of the satellite order.

As illustrated in Fig.~\ref{fig:selector_fig}, we first map sCSI feature ${\boldsymbol{\psi}}^{\rm sel}_{s,k}$ into ${\bf h}^{(0)}_{s,k}\in\mathbb{R}^{H_{\rm sel}}$ in the common hidden space using a shared MLP. 
We then complete the CSI interaction among different satellites by using a satellite-dimension Transformer encoder. Since the parameters of each layer are shared across satellites, this stage is permutation equivariant with respect to the satellite order. The corresponding expression is given by
\begin{align}
    \widetilde{\bf H}^{(\ell)}_{k}
    &\textstyle=
    {\bf H}^{(\ell-1)}_{k}
    +
    \mathrm{MHSA}(
        \mathrm{LN}\bigl(
            {\bf H}^{(\ell-1)}_{k}
            \bigr)
    ), \\
    {\bf H}^{(\ell)}_{k}
    &=
    \widetilde{\bf H}^{(\ell)}_{k}
    +
    \mathrm{FFN}\bigl(
        \mathrm{LN}\bigl(
            \widetilde{\bf H}^{(\ell)}_{k}
            \bigr)
    \bigr),
    \quad \ell=1,\ldots,N_{\rm sel}, \\
    {\bf H}^{\rm sat}_{k}
    &=
    \mathrm{LN}_{\rm sat}\bigl(
        {\bf H}^{(N_{\rm sel})}_{k}
    \bigr)
    \in
    \mathbb{R}^{S_k\times H_{\rm sel}},
    \label{eq:selector_sat_transformer}
\end{align}
where $\mathrm{FFN}(\cdot)$ denotes a feed-forward network (FFN), and $N_{\rm sel}$ denotes the depth. 

Then, we adopt set Transformer-based attention pooling to achieve permutation-invariant global feature extraction \cite{set-transformer}:
\begin{align}
    \widetilde{\bf u}_{k}
    &=
    \mathrm{MHA}\bigl(
        {\bf q}_{\rm sat},
        {\bf H}^{\rm sat}_{k},
        {\bf H}^{\rm sat}_{k}
    \bigr)
    \in
    \mathbb{R}^{1\times H_{\rm sel}}, \\
    \bar{\bf u}_{k}
    &=
    \mathrm{LN}^{\rm sat}_{1}\bigl(
        {\bf q}_{\rm sat}
        +
        \widetilde{\bf u}_{k}
    \bigr)
    \in
    \mathbb{R}^{1\times H_{\rm sel}}, \\
    {\bf u}_{k}
    &=
    \mathrm{LN}^{\rm sat}_{2}\bigl(
        \bar{\bf u}_{k}
        +
        \mathrm{FFN}^{\rm sat}_{\rm pool}\bigl(
            \bar{\bf u}_{k}
        \bigr)
    \bigr)
    \in
    \mathbb{R}^{H_{\rm sel}},
\end{align}
where $\mathrm{MHA}(\cdot)$ denotes multi-head attention, and ${\bf q}_{\rm sat}\in\mathbb{R}^{1\times H_{\rm sel}}$ is a learnable seed query. Compared with simple mean pooling, this mechanism allows the selector to place more emphasis on those satellites whose CSI are more informative for the CT/NCT selection.
After the global CSI feature has been formed, the condition vector ${\bf c}_{\rm sys}\triangleq[P_{\rm T},\lambda]^{T}\in\mathbb{R}^{2}$ is mapped into the common hidden space by a two-layer MLP, yielding ${\bf c}_{\rm hid}\in\mathbb{R}^{H_{\rm sel}}$.
The final mode logits are then produced by a lightweight classification head,
\vspace{-2mm}
\begin{align}
    &\qquad\qquad\quad{\bf t}_{k}
    =
    \mathrm{concat}\bigl(
        {\bf u}_k,
        {\bf c}_{\rm hid}
    \bigr)
    \in
    \mathbb{R}^{2H_{\rm sel}}, \\
    &{\boldsymbol{o}}_{k}
    =
    {\rm MLP}(\mathrm{LN}\bigl(
        {\bf t}_{k}
    \bigr))
    \in
    \mathbb{R}^{2},\  
    \hat m_k
    =
    \arg\max_{m\in\{0,1\}}
    [{\boldsymbol{o}}_{k}]_m,
    \label{eq:selector_decision}
\end{align}
where $\hat m_k=0$ corresponds to CT and $\hat m_k=1$ corresponds to NCT.
It is straightforward to verify that the constructed mode-switching network satisfies the permutation-invariant property for the satellite order, which stems from the stack of permutation-equivariant and permutation-invariant networks.

Finally, the selector is trained as a binary classifier using cross-entropy loss, with the label $m_k^{\star}$ generated online by the frozen CT and NCT SemCom networks during training. The loss function is given by
\vspace{-2mm}
\begin{align}
    \min_{f_{\rm sel}}
    \ \mathbb{E}
    \left[
        \mathcal{L}_{\rm CE}
        \left(
            {\boldsymbol{o}}_{k},
            m_k^{\star}
        \right)
    \right].
    \label{eq:selector_ce}
\end{align}
The switching operation illustrated in \figref{fig:ms_semcom_framework} is schematic. In practical deployment, the master satellite can select the cooperative mode and distribute the resulting mode instruction to the other cooperating satellites and the UT.

\begin{remark}
    Since the mode-switching network takes long-term multi-satellite sCSI as its input, the resulting mode decision remains valid over an extended time scale and does not require frequent updates. Its computation and signaling overheads are therefore small compared with semantic encoding and decoding, and are incurred only on a long time scale.
\end{remark}

\vspace{-2mm}
\subsection{Training Details}
\vspace{-1mm}
Each image is partitioned into $p_1\times p_2$ patches with $p_1=p_2=4$. The MSCT encoder and decoder adopt $D_1=128$ and $D_2=256$, where $N^{\rm CT}_{\rm Enc,1}=N^{\rm CT}_{\rm Enc,2}=N^{\rm CT}_{\rm Dec,1}=N^{\rm CT}_{\rm Dec,2}=4$. The window size is $w_{\rm h}\times w_{\rm w}$ with $w_{\rm h}=w_{\rm w}=4$, the number of attention heads is $N_{\rm h}=8$ in the first stage and $16$ in the second stage, and each HSTC block uses an MLP width $384$ with a bottleneck convolutional branch of reduction ratio $r_{\rm c}=4$, i.e., $D_r=D/r_{\rm c}=D/4$. The MSNCT adopts a Transformer-based backbone with $D_1=128$, FFN dimension $512$, $8$ attention heads, and encoder depth $N^{\rm NCT}_{\rm Enc}=8$; its two-stage decoder is evenly split by default as $N^{\rm NCT}_{\rm Dec,SW}=N^{\rm NCT}_{\rm Dec,CS}=4$. The mode switching network uses hidden dimension $H_{\rm sel}=128$, an $N_{\rm sel}=2$-layer satellite-wise Transformer with $4$ attention heads and FFN dimension $512$, followed by Set Transformer pooling. The tradeoff factor is set to $\lambda=0.99$ unless otherwise specified. The CT/NCT networks are optimized using Adam with a fixed learning rate of $10^{-4}$, batch size $256$, and $2000$ training epochs, while the mode selector is trained for $500$ epochs.
Specifically, in each epoch, sCSI are regenerated and random channels are generated for training, whereas the validation and test channels are kept fixed for repeatability. To enhance robustness to link-budget variations, the transmit power for every mini-batch is independently uniformly sampled between $30$ and $45$ dBm.

\begin{figure}[t]
    \centering
    \includegraphics[width=0.5\textwidth]{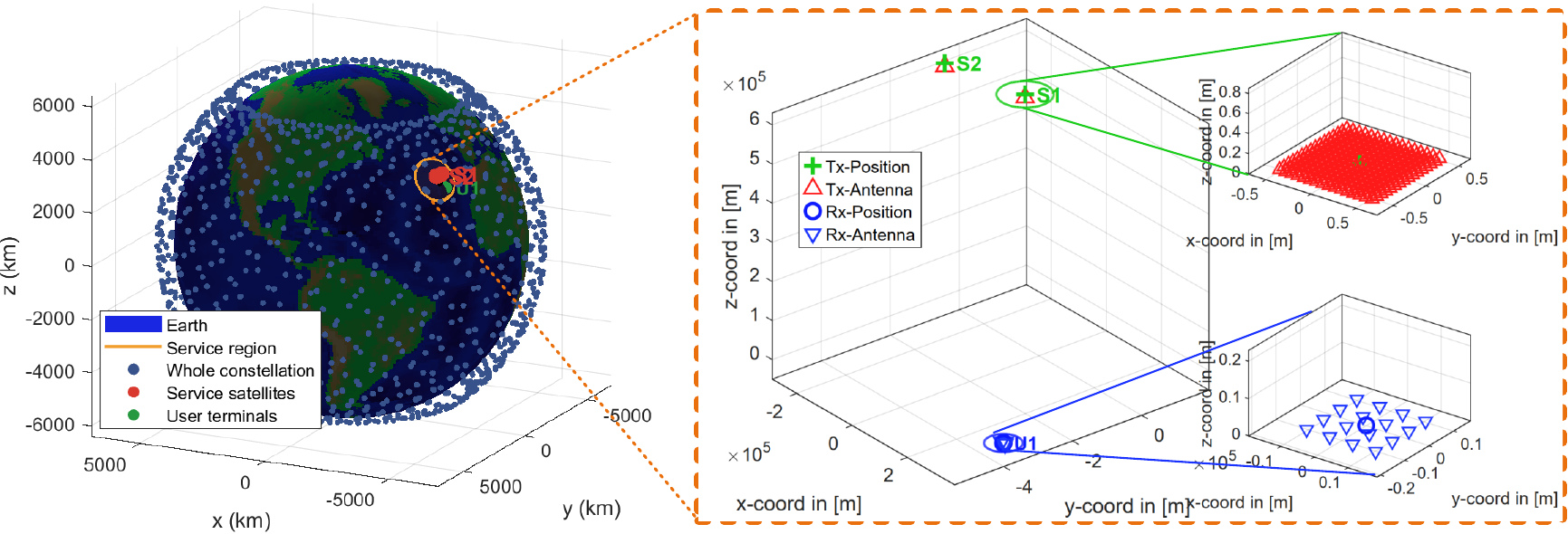}
    \caption{Visualization of one Monte Carlo realization in the simulated LEO SemCom scenario.}
    \label{fig:earth_monte_carlo_scene}
    \vspace{-5mm}
\end{figure}

\begin{table}[t]
    \centering
    \caption{Simulation parameters \cite{SpaceX_Gen2_2021,3GPP_TR_38_811,3gpp_tr_38_821,you2020massive,wang2026DP_JSAC}.}
    \label{tab:sim_satcom_setup}
    \footnotesize
    \setlength{\tabcolsep}{4pt}
    \renewcommand{\arraystretch}{1.05}
    \begin{tabular}{ll}
        \toprule
        Parameter & Value \\
        \midrule
        Constellation type & Walker-Delta \\
        Orbital altitude & $600$ km \\
        Orbital inclination & $53^\circ$ \\
        Orbital planes & $28$ \\
        Satellites per plane & $60$ \\
        Cooperative satellites & $S_k=2$ \\
        Satellite array & $N_{\rm T}=16\times 16$ \\
        Service radius & $800$ km \\
        Carrier frequency & $2.185$ GHz \\
        Diffuse rays & $12$ \\
        Satellite element gain & $6$ dBi \\
        UT element gain & $0$ dBi \\
        Noise figure & $7$ dB \\
        System bandwidth & $20$ MHz \\
        \bottomrule
    \end{tabular}
    \vspace{-5mm}
\end{table}

\vspace{-1mm}
\section{Simulation Results}
\label{sec:simulation_results}

Monte Carlo simulations are conducted in this section, where the service-region center, serving satellites, UT position, and array orientations in each sample are generated following \cite{wang2026DP_JSAC,wang2026stf}.
\figref{fig:earth_monte_carlo_scene} visualizes one random realization and provides an intuitive illustration of the simulation geometry, the cooperative-satellite selection, and the transceiver arrays.
We generate the massive MIMO channel and sCSI following the approach similar to \cite{you2020massive,li2021downlink}. The large-scale gain is determined by free-space path loss and log-normal shadowing, while the NLoS component is modeled as a complex Gaussian term whose receive-side covariance is constructed from NLoS paths. Key parameters include the Rician factor $\bar{\kappa}_{s,k}=10\log_{10}\kappa_{s,k}\in[8,15]$ dB and the number of NLoS paths $L_{s,k}-1=12$. The remaining main simulation parameters are summarized in Table~\ref{tab:sim_satcom_setup}.
To facilitate experiments while emphasizing framework innovation rather than dataset-specific performance benchmarking, CIFAR-10 is adopted for training, validation, and testing. Owing to the image-size scalability of the adopted backbones, the experiments can be readily extended to higher-resolution image datasets such as {EuroSAT} and {ImageNet} \cite{helber2019eurosat,deng2009imagenet}. Reconstruction quality is evaluated by PSNR and the structural similarity index measure (SSIM) expressed in dB, where ${\rm SSIM~(dB)}=-10\log_{10}(1-{\rm SSIM})$. Each point is averaged over $500$ test samples.

\begin{figure*}[t]
    \centering
    \subfloat[$N_{\rm R}=4\times 4$, CR$=24$.]{\includegraphics[width=0.245\textwidth]{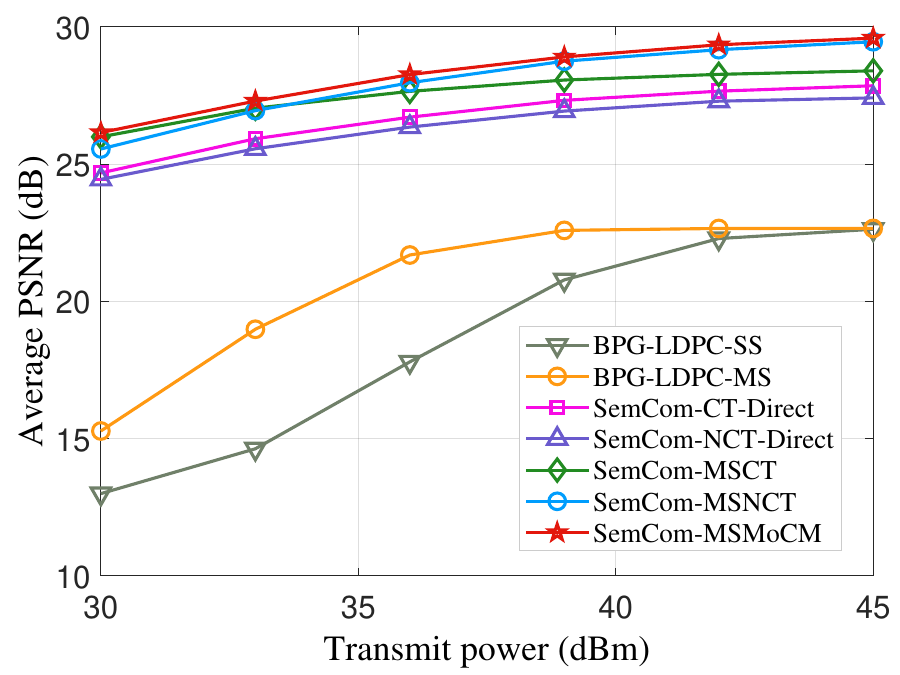}\label{fig:psnr_vs_power_R4C24}}
    \subfloat[$N_{\rm R}=4\times 4$, CR$=12$.]{\includegraphics[width=0.245\textwidth]{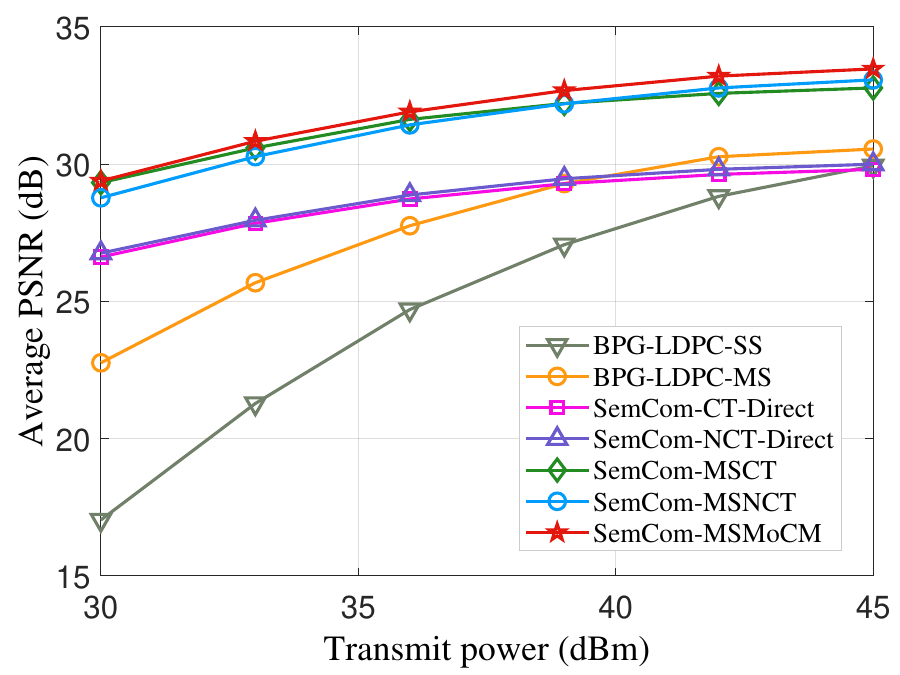}\label{fig:psnr_vs_power_R4C12}}
    \subfloat[$N_{\rm R}=2\times 2$, CR$=24$.]{\includegraphics[width=0.245\textwidth]{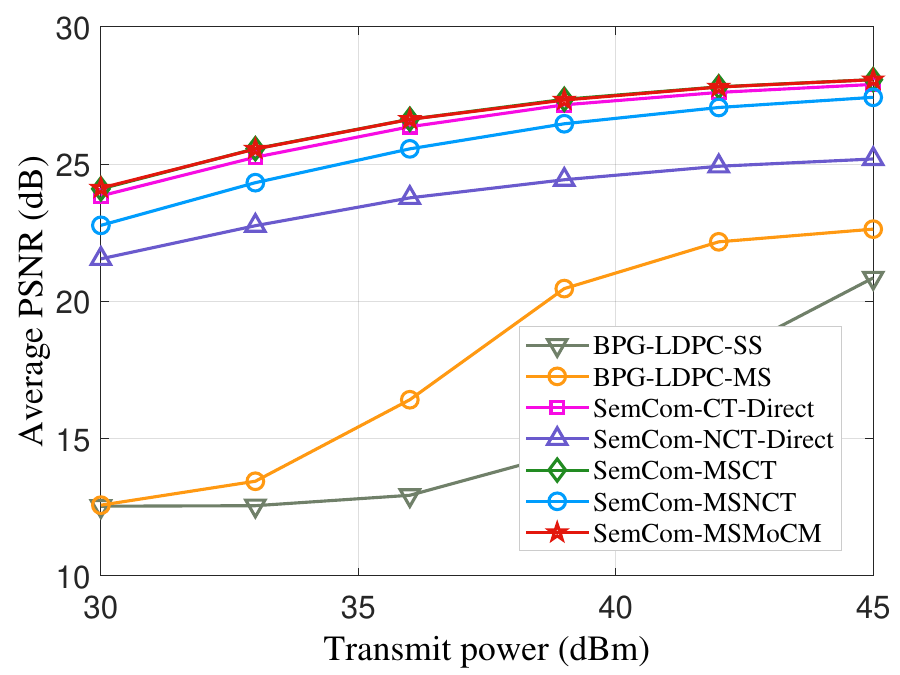}\label{fig:psnr_vs_power_R2C24}}
    \subfloat[$N_{\rm R}=2\times 2$, CR$=12$.]{\includegraphics[width=0.245\textwidth]{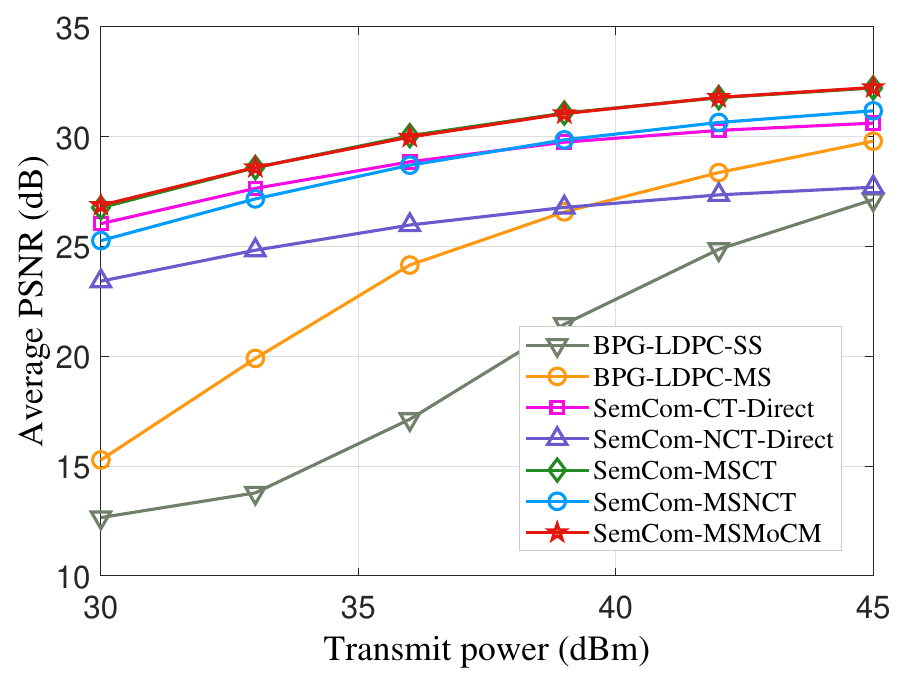}\label{fig:psnr_vs_power_R2C12}}
    \vspace{-1mm}
    \caption{Average PSNR versus transmit power under different UT array sizes and compression ratios.}
    \label{fig:psnr_vs_power_all}
    \vspace{-3mm}
\end{figure*}
\begin{figure*}[t]
    \vspace{-3mm}
    \centering
    \subfloat[$N_{\rm R}=4\times 4$, CR$=24$.]{\includegraphics[width=0.245\textwidth]{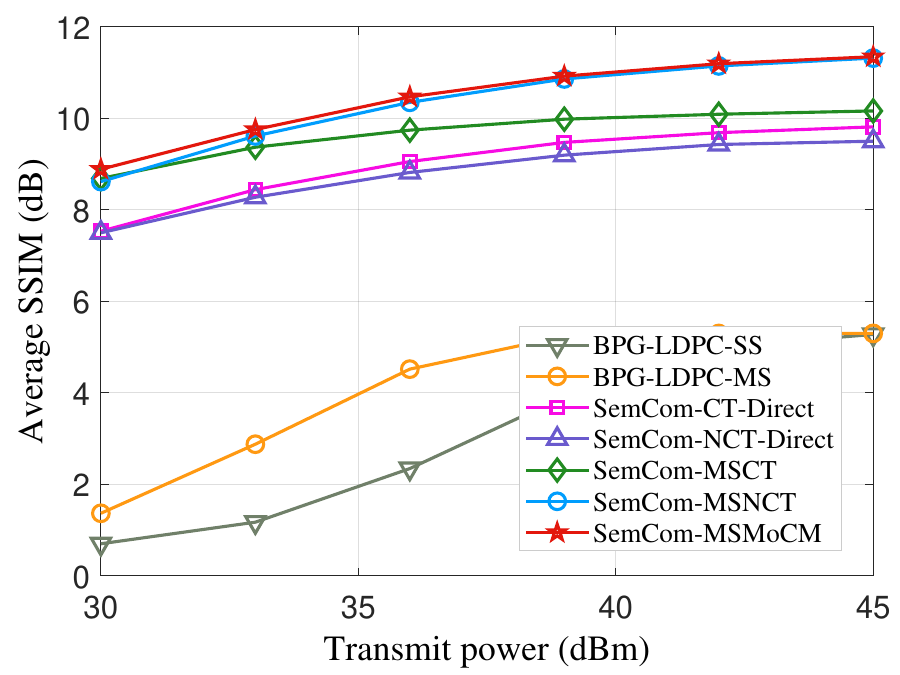}\label{fig:ssim_vs_power_R4C24}}
    \subfloat[$N_{\rm R}=4\times 4$, CR$=12$.]{\includegraphics[width=0.245\textwidth]{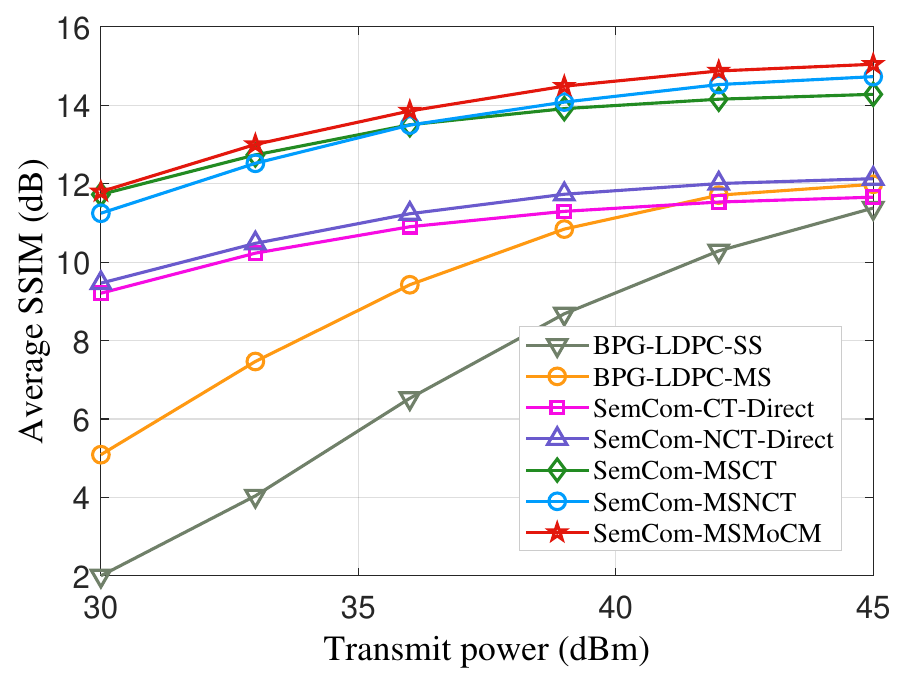}\label{fig:ssim_vs_power_R4C12}}
    \subfloat[$N_{\rm R}=2\times 2$, CR$=24$.]{\includegraphics[width=0.245\textwidth]{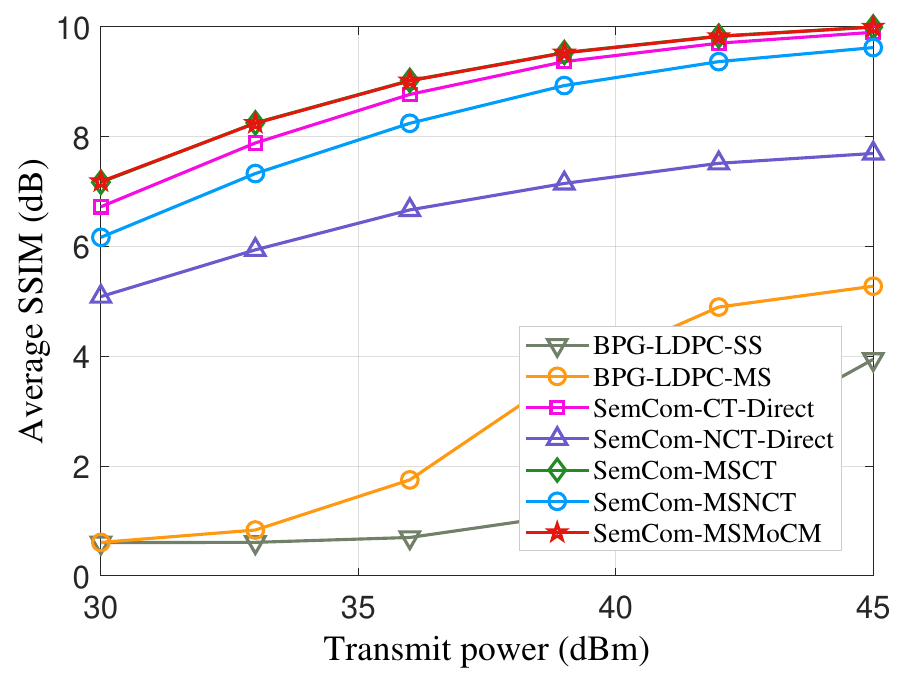}\label{fig:ssim_vs_power_R2C24}}
    \subfloat[$N_{\rm R}=2\times 2$, CR$=12$.]{\includegraphics[width=0.245\textwidth]{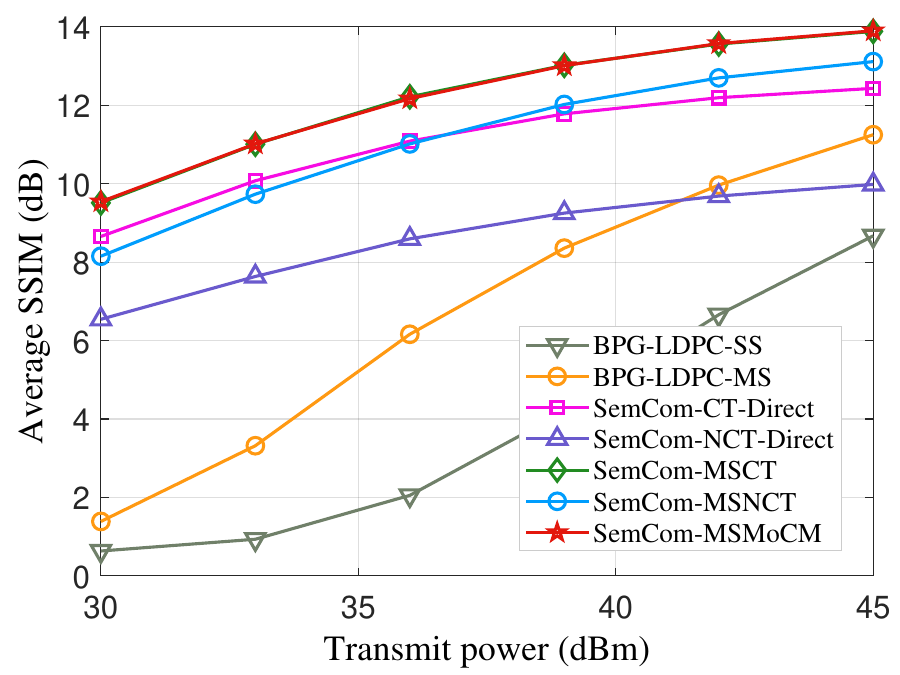}\label{fig:ssim_vs_power_R2C12}}
    \vspace{-1mm}
    \caption{Average SSIM (dB) versus transmit power under different UT array sizes and compression ratios.}
    \label{fig:ssim_vs_power_all}
    \vspace{-4mm}
\end{figure*}

To comprehensively evaluate the proposed framework, we further compare the following image transmission schemes:
\begin{itemize}
    \item \textbf{BPG-LDPC-SS}: a single-satellite baseline that cascades Better Portable Graphics (BPG) source coding with low-density parity-check (LDPC) channel coding \cite{yoo2023role}, extending the baseline in \cite{DeepJSCCMIMO2024} by allowing higher-order quadrature amplitude modulation (QAM) options.
    \item \textbf{BPG-LDPC-MS}: the MSCT version of BPG-LDPC-SS.
    \item \textbf{SemCom-CT-Direct}: a direct extension of SemCom to CT transmission in the considered scenario, where the receiver directly demodulates the semantic information from the received signal with HSTC-based backbones.
    \item \textbf{SemCom-NCT-Direct}: a direct extension of SemCom to NCT transmission in the considered scenario, in which the sub-semantic streams are transmitted and decoded independently with Transformer-based backbones.
    \item \textbf{SemCom-MSCT}/\textbf{-MSNCT}/\textbf{-MSMoCM}: our proposed CT, NCT, and hybrid-mode multi-satellite cooperative SemCom frameworks, described in Sections~\ref{framework ct sec},~\ref{framework nct sec}, and~\ref{sec:scalable_csi_module}, respectively.
\end{itemize}

Figs.~\ref{fig:psnr_vs_power_all} and~\ref{fig:ssim_vs_power_all} report the average PSNR and SSIM (dB) as functions of the transmit power under different UT array configurations and compression ratios.
Across the entire transmit-power range, array configurations, and compression ratios, the proposed cooperative semantic schemes deliver consistent and substantial gains over the conventional baselines in both PSNR and SSIM, indicating improvements in pixel-level fidelity and structural preservation. This advantage becomes more pronounced under the higher compression ratio. In addition, SemCom-MSCT and SemCom-MSNCT consistently outperform their Direct counterparts, demonstrating that the proposed architectures can effectively exploit the performance potential of multi-satellite SemCom. Moreover, there is no universally dominant choice between SemCom-MSCT and SemCom-MSNCT \cite{wang2026stf}. Under $N_{\rm R}=2 \times 2$ and ${\rm CR}=12$, CT dominates because enhancing the SNR of a single data stream has a stronger impact under a low link budget and limited receiver-side spatial resolution, whereas under $N_{\rm R}=4\times 4$ and ${\rm CR}=24$, NCT becomes more favorable because of the enhanced spatial multiplexing capability at the receiver and the more stringent compression requirement. Notably, the proposed SemCom-MSMoCM adaptively switches between CT and NCT, thereby closely predicting the optimal cooperative mode across different metrics and operating regimes. As the transmit power further increases, all curves gradually saturate, indicating that the dominant bottleneck in the high-SNR regime shifts from channel noise to the limited semantic-symbol budget.

\begin{figure*}[t]
    \centering
    \begin{minipage}[b]{0.42\textwidth}
        \centering
        \includegraphics[height=1.35in]{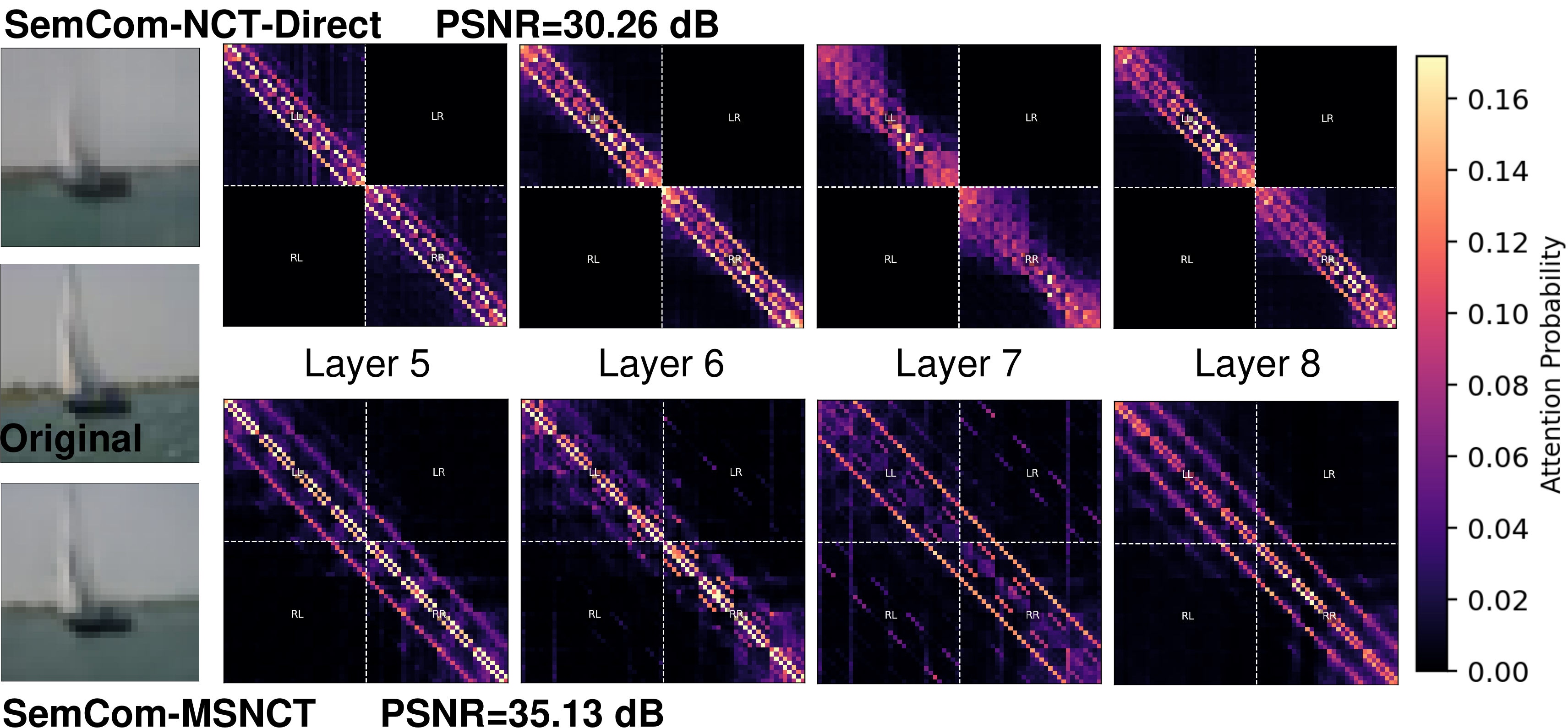}
        \vspace{-3mm}
        \caption{Reconstructed images and decoder attention maps at Layers $5$--$8$ for SemCom-NCT-Direct and SemCom-MSNCT.}
        \label{fig:nct_attention_visualization}
    \end{minipage}
    \hfill
    \begin{minipage}[b]{0.27\textwidth}
        \centering
        \includegraphics[height=1.35in]{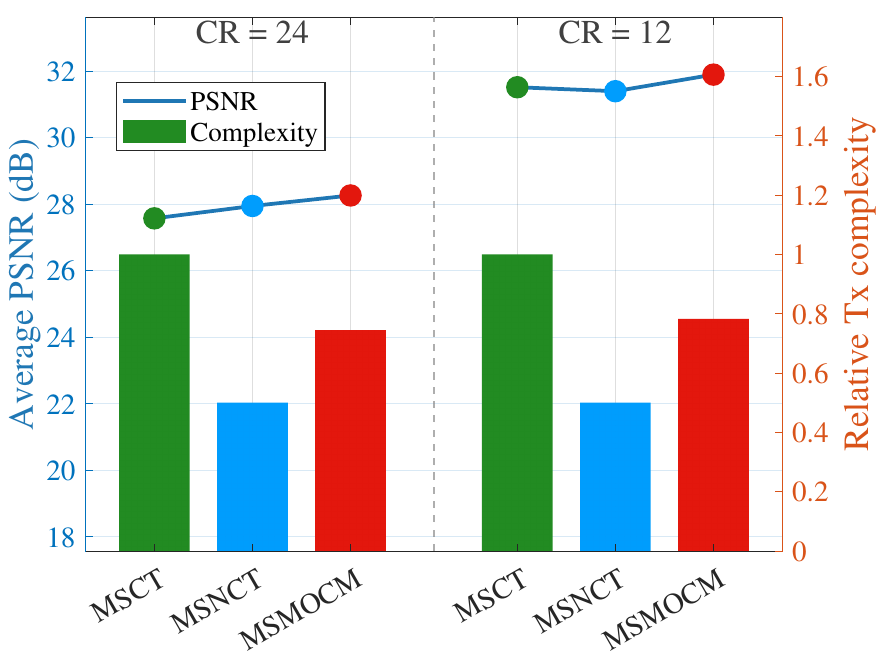}
        \vspace{-3mm}
        \caption{Performance versus relative transmitter-side complexity of the three proposed schemes under $N_\mathrm{R}=4\times 4$.}
        \label{fig:relative_complexity}
    \end{minipage}
    \hfill
    \begin{minipage}[b]{0.27\textwidth}
        \centering
        \includegraphics[height=1.35in]{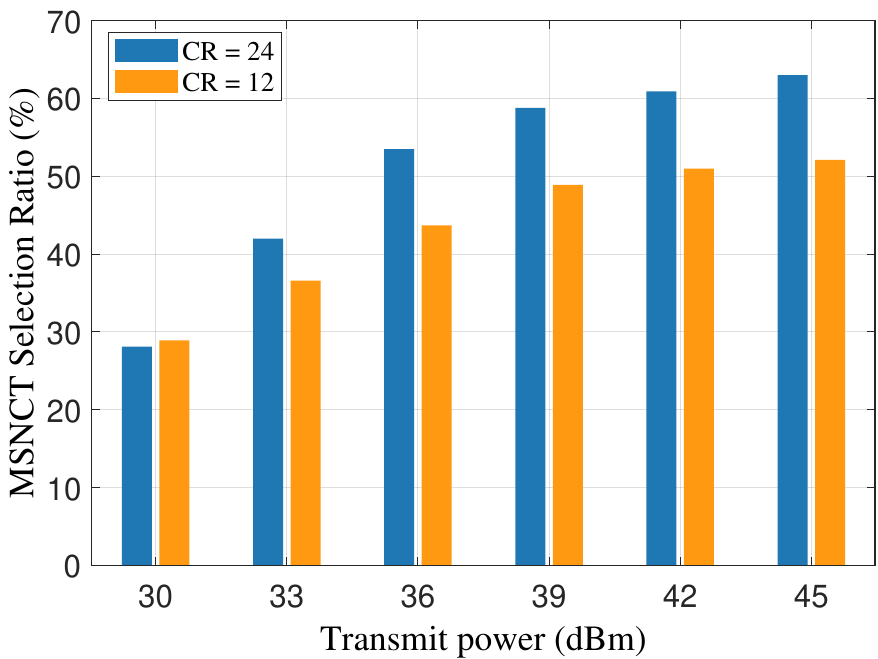}
        \vspace{-3mm}
        \caption{Proportion of the NCT mode selected by SemCom-MSMoCM versus transmit power under $N_\mathrm{R}=4\times 4$.}
        \label{fig:nct_ratio}
    \end{minipage}
    \vspace{-4mm}
\end{figure*}

\figref{fig:nct_attention_visualization} presents the reconstructed images together with the attention maps of SemCom-NCT-Direct and SemCom-MSNCT at the decoder. Since SemCom-NCT-Direct processes the two sub-semantic streams independently throughout the decoding procedure, its attention maps retain only intra-stream self-attention and therefore exhibit a block-diagonal structure. In contrast, the proposed SemCom-MSNCT aggregates the semantic streams during the cross-stream stage, leading to pronounced responses over the entire attention map. This behavior indicates effective information exchange between the two streams and reveals the exploitation of semantic interference.
This difference is also reflected in the reconstructed images. 
This observation highlights the importance of cross-stream semantic interaction through token-domain global attention.

\begin{table}[t]
    \centering
    \caption{Backbone comparison, including (a) PSNR and FLOPs and (b) FLOP orders, $N_{\rm R}=4\times 4$, ${\rm CR}=24$.}
    \label{tab:backbone_comparison}
    \vspace{-3mm}
    {\footnotesize
    \setlength{\tabcolsep}{4pt}
    \renewcommand{\arraystretch}{0.9}
    \begin{tabular*}{\columnwidth}{@{\extracolsep{\fill}}lcccc@{}}
        \multicolumn{5}{c}{\textbf{(a) Performance and FLOPs}} \\
        \toprule
        Scheme & \multicolumn{2}{c}{PSNR (dB)} & \multicolumn{2}{c}{FLOPs (MFLOPs)} \\
        \cmidrule(lr){2-3}\cmidrule(lr){4-5}
            & Test & Best Val.\,$\le$200 & TX & RX \\
        \midrule
        \textbf{MSCT-HSTC}      & \textbf{27.57} & \textbf{26.68} & 219.8 & 219.8 \\
        MSCT-HSTC-noC  & 27.48          & 26.53          & 211.1 & 211.1 \\
        MSCT-TF       & 27.49          & 25.79          & 220.9 & 233.7 \\
                \midrule
        \textbf{MSNCT-TF}      & \textbf{27.97} & 26.31          & 106.3 & 233.9 \\
        MSNCT-HSTC     & 26.76          & 25.52          & 109.7 & 223.6 \\
        \bottomrule
    \end{tabular*}}

    \vspace{1.2mm}

    {\scriptsize
    \setlength{\tabcolsep}{3pt}
    \renewcommand{\arraystretch}{1.12}
    \begin{tabular*}{\columnwidth}{@{\extracolsep{\fill}}lll@{}}
        \multicolumn{3}{c}{\textbf{(b) FLOP orders}} \\
        \toprule
        Scheme & TX order & RX order \\
        \midrule
        MSCT-HSTC$^{\dagger}$ & $\mathcal{O}(N^{\rm CT}_{\rm Enc,1}\xi_1+N^{\rm CT}_{\rm Enc,2}\xi_2)$ & $\mathcal{O}(N^{\rm CT}_{\rm Dec,2}\xi_2+N^{\rm CT}_{\rm Dec,1}\xi_1)$ \\
        MSCT-TF   & $\mathcal{O}(N^{\rm CT}_{\rm Enc}\eta)$       & $\mathcal{O}(N^{\rm CT}_{\rm Dec}\eta)$ \\
        MSNCT-TF  & $\mathcal{O}(N^{\rm NCT}_{\rm Enc}\bar{\eta})$       & $\mathcal{O}(S_kN^{\rm NCT}_{\rm Dec,SW}\bar{\eta}+N^{\rm NCT}_{\rm Dec,CS}\eta)$ \\
        MSNCT-HSTC & $\mathcal{O}(N^{\rm NCT}_{\rm Enc,1}\bar{\xi}_1+N^{\rm NCT}_{\rm Enc,2}\bar{\xi}_2)$ & $\mathcal{O}(S_kN^{\rm NCT}_{\rm Dec,SW}\bar{\xi}_2+N^{\rm NCT}_{\rm Dec,CS}\xi_1)$ \\
        \bottomrule
    \end{tabular*}}

    \vspace{0.6mm}
    \parbox{\columnwidth}{\scriptsize
    $^{\dagger}$ Similar for MSCT-HSTC-noC.\;
    Auxiliary symbols:
    $\xi_i\triangleq P_iD_i^2+P_iP_{\rm w}D_i$,
    $\eta\triangleq P_1D_1^2+P_1^2D_1$,
    $\bar{\eta}\triangleq P^{\rm sub}D_1^2+(P^{\rm sub})^2D_1$,
    $\bar{\xi}_i\triangleq P_i^{\rm sub}D_i^2+P_i^{\rm sub}P_{\rm w}D_i,\ i\in\{1,2\}$,
    $P_1^{\rm sub}\triangleq P^{\rm sub},\ P_2^{\rm sub}\triangleq P^{\rm sub}/4$.}
    \vspace{-8mm}
\end{table}

Fig.~\ref{fig:relative_complexity} compares the average PSNR of SemCom-MSCT, SemCom-MSNCT, and SemCom-MSMoCM against their relative transmitter-side burden. Since transmitter-side complexity is multi-faceted, we use the per-satellite source-data acquisition/processing amount as a representative burden metric, which yields nearly the same CT-to-NCT ratio as transmitter-side FLOPs under the considered configuration. With MSCT normalized to one, MSNCT has relative burden $1/S_k=1/2$, while SemCom-MSMoCM achieves a favorable performance-complexity tradeoff at an intermediate burden.
Fig.~\ref{fig:nct_ratio} shows the proportion of the NCT mode selected by SemCom-MSMoCM as the transmit power varies. Under the more aggressive compression setting with ${\rm CR}=24$, SemCom-MSMoCM tends to select the NCT mode with a higher probability across all power levels, whereas the overall NCT proportion decreases under ${\rm CR}=12$. As the transmit power increases, the impact of the compression ratio becomes more dominant, further raising the NCT selection proportion. These observations demonstrate that the proposed MoCM architecture can dynamically switch between its sub-architectures according to the conditions.

Table~\ref{tab:backbone_comparison} corroborates our motivation for pairing CT with HSTC and NCT with Transformer-based backbone (denoted by TF). ``MSCT-HSTC-noC'' is a zero-convolution ablation of HSTC that zeros the lightweight bottleneck CNN-side output to isolate the local inductive bias. 
``Test'' denotes the test PSNR performance averaged over the considered transmit power range, ``Best Val.\,$\le$200'' reports the best validation PSNR within the first $200$ epochs as a proxy for early-stage convergence, and the TX/RX FLOPs are the per-image forward-pass costs at the satellite and UT, respectively. Under CT, HSTC attains the most favorable performance-complexity tradeoff comparable to other backbones. Under NCT, TF prevails owing to the natural fit between its global self-attention and cross-stream token-level interaction, producing a backbone ranking opposite to that on the CT side. These results validate the proposed backbone designs under different cooperative architectures.

\vspace{-2mm}
\section{Conclusion}
\label{sec:conclusion}
\vspace{-1mm}
This paper investigated SemComs for multi-satellite cooperative massive MIMO transmission, developing tailored frameworks for both the coherent and non-coherent transmission modes. The MSCT design adopted a symmetric HSTC-based encoder-decoder for scalable coherent semantic reconstruction, whereas the MSNCT design combined transmitter-side stream allocation with a two-stage Transformer-based receiver to exploit cross-stream semantic interference. Building upon them, an MoCM framework was further proposed, in which a permutation-invariant network leverages multi-satellite sCSI to adaptively switch between the two modes. Simulations under practical LEO settings confirmed consistent reconstruction gains across the proposed frameworks and showed that MoCM delivers a favorable performance--complexity tradeoff, taking an initial step toward the deep integration of multi-satellite cooperative transmission and SemComs.

\bibliographystyle{IEEEtran}
\bibliography{reference}

@IEEEtranBSTCTL{IEEEexample:BSTcontrol,
  CTLuse_article_number = "yes",
  CTLuse_paper = "yes",
  CTLuse_forced_etal = "yes",
  CTLmax_names_forced_etal = "4",
  CTLnames_show_etal = "4",
  CTLuse_alt_spacing = "yes",
  CTLalt_stretch_factor = "4",
  CTLdash_repeated_names = "yes",
  CTLname_format_string = "{f.~}{vv~}{ll}{, jj}",
  CTLname_latex_cmd = "",
  CTLname_url_prefix = "[Online]. Available:"
 }

@Article{wu2024large,
  author    = {Wu, Yezeng and Xiao, Lixia and Zhou, Jiaxi and Feng, Mingjie and Xiao, Pei and Jiang, Tao},
  journal   = {IEEE Commun. Mag.},
  title     = {Large-Scale {MIMO} Enabled Satellite Communications: Concepts, Technologies, and Challenges},
  year      = {2024},
}

@article{you2020massive,
  author  = {Li You and Ke-Xin Li and Jiaheng Wang and Xiqi Gao and Xiang-Gen Xia and Bj{\"o}rn Ottersten},
  title   = {Massive {MIMO} transmission for {LEO} satellite communications},
  journal = {IEEE J. Sel. Areas Commun.},
  year    = {2020},
  volume  = {38},
  number  = {8},
  pages   = {1851--1865},
  month   = {Aug.}
}

@Article{you2022hybrid,
  author    = {You, Li and Qiang, Xiaoyu and Li, Ke-Xin and Tsinos, Christos G and Wang, Wenjin and Gao, Xiqi and Ottersten, Bj{\"o}rn},
  journal   = {IEEE Trans. Wireless Commun.},
  title     = {Hybrid analog/digital precoding for downlink massive {MIMO} {LEO} satellite communications},
  year      = {2022},
  number    = {8},
  pages     = {5962--5976},
  volume    = {21},
}

@Article{9193995,
  author   = {Kibria, Mirza Golam and Lagunas, Eva and Maturo, Nicola and Al-Hraishawi, Hayder and Chatzinotas, Symeon},
  journal  = {IEEE Open J. Commun. Soc.},
  title    = {Carrier Aggregation in Satellite Communications: Impact and Performance Study},
  year     = {2020},
  pages    = {1390--1402},
  volume   = {1},
  doi      = {10.1109/OJCOMS.2020.3023375},
}

@article{xiang2024massive,
  author  = {Xiang, Ziyu and Gao, Xiqi and Li, Ke-Xin and Xia, Xiang-Gen},
  title   = {Massive {MIMO} Downlink Transmission for Multiple {LEO} Satellite Communication},
  journal = {IEEE Trans. Commun.},
  year    = {2024},
  volume  = {72},
  number  = {6},
  pages   = {3352--3364},
  month   = {Jun.}
}

@ARTICLE{li2021downlink,
  author  = {K.-X. Li and L. You and J. Wang and X. Gao and C. G. Tsinos and S. Chatzinotas and B. Ottersten},
  title   = {Downlink Transmit Design for Massive {MIMO} {LEO} Satellite Communications},
  journal = {IEEE Trans. Commun.},
  volume  = {70},
  number  = {2},
  pages   = {1014--1028},
  month   = {Feb.},
  year    = {2021},
  doi     = {10.1109/TCOMM.2021.3131573}
}

@article{wang2026DP_JSAC,
  author  = {Wang, Yafei and Ha, Vu Nguyen and Ntontin, Konstantinos and Yan, Hong and Wang, Wenjin and Chatzinotas, Symeon and Ottersten, Bj{\"o}rn},
  title   = {Statistical {CSI}-Based Distributed Precoding Design for {OFDM}-Cooperative Multi-Satellite Systems},
  journal = {IEEE J. Sel. Areas Commun.},
  year    = {2026},
  volume  = {44},
  pages   = {3219--3236},
  month   = {Jan.},
  doi     = {10.1109/JSAC.2026.3650895}
}

@ARTICLE{hou2024joint,
  author  = {H. Hou and Y. Wang and X. Yi and W. Wang and S. Jin},
  title   = {Joint Beam Alignment and Doppler Estimation for Fast Time-Varying Wideband {mmWave} Channels},
  journal = {IEEE Trans. Wireless Commun.},
  volume  = {23},
  number  = {9},
  pages   = {10895--10910},
  month   = {Sep.},
  year    = {2024}
}

@techreport{3GPP_TR_38_811,
  author       = {3GPP},
  title        = {{TR} 38.811 V15.4.0: Study on New Radio ({NR}) to Support Non-Terrestrial Networks},
  institution  = {3GPP},
  type         = {Tech. Rep.},
  number       = {TR 38.811 V15.4.0},
  year         = {2020},
  month        = {Sep.}
}

@techreport{3gpp_tr_38_821,
  author       = {3GPP},
  title        = {{TR} 38.821 V16.2.0: Solutions for {NR} to Support Non-Terrestrial Networks ({NTN})},
  institution  = {3GPP},
  type         = {Tech. Rep.},
  number       = {TR 38.821 V16.2.0},
  year         = {2023},
  month        = {Mar.}
}

@techreport{SpaceX_Gen2_2021,
  title        = {Federal Communications Commission; Amendment to Pending Application for the {SpaceX} {Gen2} {NGSO} Satellite System},
  institution  = {FCC},
  number       = {File No. SAT-AMD-2021},
  type         = {Tech. Rep.},
  month        = {August},
  year         = {2021},
  address      = {Washington, D.C.},
  note         = {Available: https://fcc.report/IBFS/SAT-AMD-20210818-00105/12943361.pdf}
}

@misc{ASTBlueWalker3,
  author = {{AST SpaceMobile}},
  title  = {BlueWalker 3},
  year   = {2026},
  note   = {Available: \url{https://ast-science.com/spacemobile-network/bluewalker-3/}. Accessed: Mar. 27, 2026}
}

@article{10820534,
  author  = {Konstantinos Ntontin and Eva Lagunas and Jorge Querol and Junaid ur Rehman and Joel Grotz and Symeon Chatzinotas and Bj{\"o}rn Ottersten},
  title   = {A Vision, Survey, and Roadmap Toward Space Communications in the {6G} and Beyond Era},
  journal = {Proc. IEEE},
  volume  = {},
  number  = {},
  pages   = {1--37},
  month   = {Jan.},
  year    = {2025},
  doi     = {10.1109/JPROC.2024.3512934}
}

@article{WANG2025,
title = {Toward Mobile Satellite Internet: The Fundamental Limitation of Wireless Transmission and Enabling Technologies},
journal = {Engineering},
year = {2025},
issn = {2095-8099},
doi = {https://doi.org/10.1016/j.eng.2025.07.007},
url = {https://www.sciencedirect.com/science/article/pii/S2095809925003698},
author = {Wenjin Wang and Yiming Zhu and Yafei Wang and Rui Ding and Symeon Chatzinotas}
}

@article{11049893,
  author    = {Wang, Yafei and Hou, Hongwei and Yi, Xinping and Wang, Wenjin and Jin, Shi},
  title     = {Toward Unified {AI} Models for {MU-MIMO} Communications: A Tensor Equivariance Framework},
  journal   = {IEEE Trans. Wireless Commun.},
  year      = {2025},
  volume    = {24},
  number    = {12},
  pages     = {10517--10533},
  month     = {Dec.},
  doi       = {10.1109/TWC.2025.3580321}
}

@Article{wang2022weighted,
  author  = {Wang, Yafei and Wang, Wenjin and You, Li and Tsinos, Christos G. and Jin, Shi},
  journal = {IEEE Wireless Commun. Lett.},
  title   = {Weighted {MMSE} Precoding for Constructive Interference Region},
  year    = {2022},
  number  = {12},
  pages   = {2605--2609},
  volume  = {11},
  doi     = {10.1109/LWC.2022.3211731},
}

@inproceedings{10437228,
  author    = {Shiyu Wu and Yafei Wang and Gangle Sun and Li You and Wenjin Wang and Rui Ding},
  title     = {Energy and Computational Efficient Precoding for {LEO} Satellite Communications},
  booktitle = {Proc. IEEE Glob. Commun. Conf. ({GLOBECOM})},
  address   = {Kuala Lumpur, Malaysia},
  month     = {Dec.},
  year      = {2023},
  pages     = {1872--1877},
  doi       = {10.1109/GLOBECOM54140.2023.10437228}
}

@article{wu2025distributed,
  title        = {Distributed Beamforming for Multiple {LEO} Satellites With Imperfect Delay and {Doppler} Compensations: Modeling and Rate Analysis},
  author       = {Wu, Shiyu and Wang, Yafei and Sun, Gangle and Wang, Wenjin and Wang, Jiaheng and Ottersten, Bj{\"o}rn},
  journal      = {IEEE Trans. Veh. Technol.},
  year         = {2025},
  volume    = {74},
  number    = {9},
  pages     = {14978--14984},
  month     = {Sep.},
  doi       = {10.1109/TVT.2025.3564047},
}

@INPROCEEDINGS{ha2024user,
  author    = {V. N. Ha and D. H. N. Nguyen and J. C.-M. Duncan and J. L. Gonzalez-Rios 
               and J. A. V. Peralvo and G. Eappen and L. M. Garces-Socarras 
               and R. Palisetty and S. Chatzinotas and B. Ottersten},
  title     = {User-Centric Beam Selection and Precoding Design for Coordinated Multiple-Satellite Systems},
  booktitle = {Proc. IEEE 35th Int. Symp. Pers., Indoor Mobile Radio Commun. ({PIMRC})},
  address   = {Valencia, Spain},
  month     = {Sep.},
  year      = {2024},
  pages     = {1--6}
}

@article{zhu2024joint,
  author    = {Zhu, Yiming and Zhuang, Jiawei and Sun, Gangle and Hou, Hongwei and You, Li and Wang, Wenjin},
  title     = {Joint Channel Estimation and Prediction for Massive {MIMO} With Frequency Hopping Sounding},
  journal   = {IEEE Trans. Commun.},
  year      = {2025},
  volume    = {73},
  number    = {7},
  pages     = {5139--5154},
  month     = {Jul.},
  doi       = {10.1109/TCOMM.2024.3523972}
}

@article{zhang2025decentralized,
  author  = {Zhang, Yuchen and Lagunas, Eva and Zheng, Xue Xian and Chatzinotas, Symeon and Al-Naffouri, Tareq Y.},
  title   = {Decentralized Cooperative Beamforming for Networked {LEO} Satellites with Statistical {CSI}},
  journal = {arXiv preprint arXiv:2512.18890},
  year    = {2025}
}

@article{Bakhsh2024MultiSatSurvey,
  author  = {Zohre Mashayekh Bakhsh and Yasaman Omid and Gaojie Chen and Farbod Kayhan and Yi Ma and Rahim Tafazolli},
  title   = {Multi-Satellite {MIMO} Systems for Direct Satellite-to-Device Communications: A Survey},
  journal = {IEEE Commun. Surveys Tuts.},
  year    = {2025},
  volume  = {27},
  number  = {3},
  pages   = {1536--1564},
  month   = {Jun.},
  doi     = {10.1109/COMST.2024.3449430}
}

@article{Xie2021DeepSC,
  author  = {Huiqiang Xie and Zhijin Qin and Geoffrey Ye Li and Biing-Hwang Juang},
  title   = {Deep Learning Enabled Semantic Communication Systems},
  journal = {IEEE Trans. Signal Process.},
  year    = {2021},
  volume  = {69},
  pages   = {2663--2675},
  doi     = {10.1109/TSP.2021.3071210}
}

@article{SemanticCommSurvey,
  author  = {Tilahun M. Getu and Georges Kaddoum and Mehdi Bennis},
  title   = {Semantic Communication: A Survey on Research Landscape, Challenges, and Future Directions},
  journal = {Proc. IEEE},
  year    = {2024},
  volume  = {112},
  number  = {11},
  pages   = {1649--1685},
  month   = {Nov.},
  doi     = {10.1109/JPROC.2024.3520707}
}

@article{Lin2026SatelliteSemCom,
  author  = {Zhongze Lin and Hui Lin and Yao Sun and Shakila Basheer and Mohammad Tabrez Quasim and Kapal Dev},
  title   = {Joint Coding and Modulation for Robust Semantic Communication in Satellite Communications},
  journal = {IEEE Internet Things J.},
  year    = {2026},
  volume  = {13},
  number  = {1},
  pages   = {339--346},
  month   = {Jan.},
  doi     = {10.1109/JIOT.2025.3616257}
}

@article{wang2026stf,
  author  = {Yafei Wang and Xiaofan Xu and Yiming Zhu and Wenjin Wang and Rui Ding and Symeon Chatzinotas and Bj{\"o}rn Ottersten},
  title   = {Multi-{LEO} Satellite Cooperative Transmission: A Spatial-Temporal-Frequency Perspective},
  journal = {IEEE Wireless Commun.},
  volume  = {},
  number  = {},
  year    = {2026},
  pages   = {1--8},
  doi     = {10.1109/MWC.2026.3659842}
}

@article{cao2026DLMSCT,
  author  = {Cao, Wenjing and Wang, Yafei and Zhang, Jinshuo and Xu, Xiaofan and Wang, Wenjin and Chatzinotas, Symeon and Ottersten, Bj{\"o}rn},
  title   = {Deep Learning-Based Multi-Satellite Massive {MIMO} Transmission: Centralized or Decentralized?},
  journal = {arXiv preprint arXiv:2603.20862},
  year    = {2026}
}

@article{11449148,
  author  = {Liu, Zijun and Wang, Yafei and Wang, Wenjin and Sun, Yi and Yan, Hong and Sun, Zhili},
  title   = {Multi-Satellite Coordinated Beam Hopping for Interference Mitigation Under Tilted Beam Effects: A Graph-Theoretic Approach},
  journal = {IEEE Wireless Commun. Lett.},
  year    = {2026},
  volume  = {15},
  pages   = {2313--2317},
  doi     = {10.1109/LWC.2026.3676112}
}

@article{Zhang2024CoopImageSemCom,
  author  = {Wenjing Zhang and Yining Wang and Mingzhe Chen and Tao Luo and Dusit Niyato},
  title   = {Optimization of Image Transmission in Cooperative Semantic Communication Networks},
  journal = {IEEE Trans. Wireless Commun.},
  year    = {2024},
  volume  = {23},
  number  = {2},
  pages   = {861--877},
  doi     = {10.1109/TWC.2023.3282906}
}

@article{DeepMA2024,
  author  = {Wenyu Zhang and Kaiyuan Bai and Sherali Zeadally and Haijun Zhang and Hua Shao and Hui Ma and Victor C. M. Leung},
  title   = {DeepMA: End-to-End Deep Multiple Access for Wireless Image Transmission in Semantic Communication},
  journal = {IEEE Trans. Cogn. Commun. Netw.},
  year    = {2024},
  volume  = {10},
  number  = {2},
  pages   = {387--402},
  doi     = {10.1109/TCCN.2023.3326302}
}

@article{DeepJSCCMIMO2024,
  author  = {Haotian Wu and Yulin Shao and Chenghong Bian and Krystian Mikolajczyk and Deniz G{\"u}nd{\"u}z},
  title   = {Deep Joint Source-Channel Coding for Adaptive Image Transmission Over {MIMO} Channels},
  journal = {IEEE Trans. Wireless Commun.},
  year    = {2024},
  volume  = {23},
  number  = {11},
  pages   = {15002--15017},
  doi     = {10.1109/TWC.2024.3427954}
}

@inproceedings{JSCCSatGroundSemCom,
  author    = {Yanbo Yin and Shu Liu and Dingzhu Wen and Youlong Wu and Yuanming Shi},
  title     = {Joint Source and Channel Coding for Multi-Modal Satellite-to-Ground Semantic Communications},
  booktitle = {Proc. IEEE Wireless Commun. Networking Conf. ({WCNC})},
  year      = {2025},
  pages     = {1--6},
  doi       = {10.1109/WCNC61545.2025.10978250}
}

@article{wang2025MSMS,
  author  = {Wang, Yafei and Zhu, Yiming and Ha, Vu Nguyen and Wang, Wenjin and Ding, Rui and Chatzinotas, Symeon and Ottersten, Bj{\"o}rn},
  title   = {Multi-Satellite Multi-Stream Beamspace Massive {MIMO} Transmission},
  journal = {arXiv preprint arXiv:2512.21998},
  year    = {2025}
}

@article{Liu2026SatAVSemCom,
  author  = {Liu, Fangyu and Jiang, Peiwen and Wang, Wenjin and Wen, Chao-Kai and Li, Xiao and Jin, Shi},
  title   = {Semantic Satellite Communications for Synchronized Audiovisual Reconstruction},
  journal = {arXiv preprint arXiv:2603.10791},
  year    = {2026}
}

@article{Bui2025LEOEO,
  author  = {Van-Phuc Bui and Thinh Quang Dinh and Israel Leyva-Mayorga and Shashi Raj Pandey and Eva Lagunas and Petar Popovski},
  title   = {Semantic Image Encoding and Communication for Earth Observation With {LEO} Satellites},
  journal = {IEEE Trans. Cogn. Commun. Netw.},
  year    = {2025},
  volume  = {11},
  number  = {2},
  pages   = {1210--1224},
  doi     = {10.1109/TCCN.2024.3451724}
}

@article{zhu2026multisatellitecooperative,
  author  = {Zhu, Yiming and Wang, Yafei and Amatetti, Carla and Vanelli-Coralli, Alessandro and Wang, Wenjin and Ding, Rui and Chatzinotas, Symeon and Ottersten, Bj{\"o}rn},
  title   = {Toward Multi-Satellite Cooperative Transmission: A Joint Framework for {CSI} Acquisition, Feedback, and Phase Synchronization},
  journal = {arXiv preprint arXiv:2603.28195},
  year    = {2026}
}

@inproceedings{swin-transformer,
  author    = {Ze Liu and Yutong Lin and Yue Cao and Han Hu and Yixuan Wei and Zheng Zhang and Stephen Lin and Baining Guo},
  title     = {Swin Transformer: Hierarchical Vision Transformer Using Shifted Windows},
  booktitle = {Proc. IEEE/CVF Int. Conf. Comput. Vis. ({ICCV})},
  year      = {2021},
  month     = {Oct.},
  pages     = {10012--10022},
  doi       = {10.1109/ICCV48922.2021.00986}
}

@inproceedings{park2022how,
  author    = {Park, Namuk and Kim, Songkuk},
  title     = {How Do Vision Transformers Work?},
  booktitle = {Proc. Int. Conf. Learn. Represent. ({ICLR})},
  year      = {2022},
  month     = {Apr.}
}

@inproceedings{vit,
  author    = {Alexey Dosovitskiy and Lucas Beyer and Alexander Kolesnikov and Dirk Weissenborn and Xiaohua Zhai and Thomas Unterthiner and Mostafa Dehghani and Matthias Minderer and Georg Heigold and Sylvain Gelly and Jakob Uszkoreit and Neil Houlsby},
  title     = {An Image is Worth 16x16 Words: Transformers for Image Recognition at Scale},
  booktitle = {Proc. Int. Conf. Learn. Represent. ({ICLR})},
  year      = {2021},
  month     = {May}
}

@inproceedings{liu2023tcm,
  author    = {Liu, Jinming and Sun, Heming and Katto, Jiro},
  title     = {Learned Image Compression with Mixed Transformer-{CNN} Architectures},
  booktitle = {Proc. IEEE/CVF Conf. Comput. Vis. Pattern Recognit. ({CVPR})},
  year      = {2023},
  pages     = {14388--14397},
  month     = {Jun.},
  doi       = {10.1109/CVPR52729.2023.01383}
}

@InProceedings{set-transformer,
  author    = {Lee, Juho and Lee, Yoonho and Kim, Jungtaek and Kosiorek, Adam R. and Choi, Seungjin and Teh, Yee Whye},
  title     = {Set Transformer: A Framework for Attention-Based Permutation-Invariant Neural Networks},
  booktitle = {Proc. 36th Int. Conf. Mach. Learn. ({ICML})},
  year      = {2019},
  month     = {Jun.},
  volume    = {97},
  pages     = {3744--3753}
}

@article{zhang2026enabling,
  author  = {Zhang, Y. and Al-Naffouri, T. Y.},
  title   = {Enabling Scalable Distributed Beamforming via Networked {LEO} Satellites Toward {6G}},
  journal = {IEEE Trans. Wireless Commun.},
  year    = {2026},
  volume  = {25},
  pages   = {6666--6680}
}

@inproceedings{deng2009imagenet,
  author    = {Deng, Jia and Dong, Wei and Socher, Richard and Li, Li-Jia and Li, Kai and Fei-Fei, Li},
  title     = {{ImageNet}: A Large-Scale Hierarchical Image Database},
  booktitle = {Proc. IEEE Conf. Comput. Vis. Pattern Recognit. ({CVPR})},
  year      = {2009},
  pages     = {248--255},
  doi       = {10.1109/CVPR.2009.5206848}
}

@article{helber2019eurosat,
  author  = {Helber, Patrick and Bischke, Benjamin and Dengel, Andreas and Borth, Damian},
  title   = {{EuroSAT}: A Novel Dataset and Deep Learning Benchmark for Land Use and Land Cover Classification},
  journal = {IEEE J. Sel. Topics Appl. Earth Observ. Remote Sens.},
  year    = {2019},
  volume  = {12},
  number  = {7},
  pages   = {2217--2226},
  doi     = {10.1109/JSTARS.2019.2918242}
}

@article{wang2019near,
  author  = {Wang, Wenjin and Tong, Yushan and Li, Lingxuan and Lu, An-An and You, Li and Gao, Xiqi},
  title   = {Near Optimal Timing and Frequency Offset Estimation for {5G} Integrated {LEO} Satellite Communication System},
  journal = {IEEE Access},
  year    = {2019},
  volume  = {7},
  pages   = {113298--113310},
  doi     = {10.1109/ACCESS.2019.2935038}
}

@article{marrero2022architectures,
  author  = {Marrero, Liz Mart{\'\i}nez and Merlano-Duncan, Juan Carlos and Querol, Jorge and Kumar, Sumit and Krivochiza, Jevgenij and Sharma, Shree Krishna and Chatzinotas, Symeon and Camps, Adriano and Ottersten, Bj{\"o}rn},
  title   = {Architectures and Synchronization Techniques for Distributed Satellite Systems: {A} Survey},
  journal = {IEEE Access},
  year    = {2022},
  volume  = {10},
  pages   = {45375--45409},
  doi     = {10.1109/ACCESS.2022.3169499}
}

@inproceedings{du2022glam,
  author    = {Du, Nan and Huang, Yanping and Dai, Andrew M. and Tong, Simon and Lepikhin, Dmitry and Xu, Yuanzhong and Krikun, Maxim and Zhou, Yanqi and Yu, Adams Wei and Firat, Orhan and Zoph, Barret and Fedus, Liam and Bosma, Maarten P. and Zhou, Zongwei and Wang, Tao and Wang, Emma and Webster, Kellie and Pellat, Marie and Robinson, Kevin and Meier-Hellstern, Kathleen and Duke, Toju and Dixon, Lucas and Zhang, Kun and Le, Quoc V. and Wu, Yonghui and Chen, Zhifeng and Cui, Claire},
  title     = {{GLaM}: Efficient Scaling of Language Models with Mixture-of-Experts},
  booktitle = {Proc. 39th Int. Conf. Mach. Learn. ({ICML})},
  year      = {2022},
  volume    = {162},
  pages     = {5547--5569}
}

@article{yoo2023role,
  author  = {Yoo, Hanju and Dai, Linglong and Kim, Songkuk and Chae, Chan-Byoung},
  title   = {On the Role of {ViT} and {CNN} in Semantic Communications: Analysis and Prototype Validation},
  journal = {IEEE Access},
  year    = {2023},
  month   = {Jul.},
  volume  = {11},
  pages   = {71528--71541},
  doi     = {10.1109/ACCESS.2023.3291405}
}
	
\end{document}